\documentclass[prd,nofootinbib,preprint,superscriptaddress]{revtex4}
\pdfoutput=1
\usepackage[utf8]{inputenc}
\usepackage{subfigure}
\usepackage{epsfig} 
\usepackage{soul} 
\usepackage{amsmath} 
\usepackage{amsfonts}  
\usepackage{float}    
\usepackage{amssymb}   
\usepackage{slashed} 
\usepackage{color}       
\usepackage{pbox}
\usepackage{array,multirow,makecell}
\usepackage[colorlinks,citecolor=blue, linkcolor = blue, urlcolor = blue]{hyperref}
\usepackage{tabularx}
\usepackage{titlesec} 
\usepackage{scrextend}
\usepackage{ulem}
\usepackage{natbib} 
\usepackage{array,multirow,makecell}
\usepackage{lipsum}

\newcommand{\lsim}[1]{\lesssim}
\begin{document}
\title{CMB signature of non-thermal Dark Matter produced from self-interacting dark sector}
\author{Dilip Kumar Ghosh}
\email{tpdkg@iacs.res.in}
\affiliation{School of Physical Sciences, Indian Association for the Cultivation of Science,\\ 2A $\&$ 2B Raja S.C. Mullick Road, Kolkata 700032, India}

\author{Purusottam Ghosh}
\email{spspg2655@iacs.res.in}
\affiliation{School of Physical Sciences, Indian Association for the Cultivation of Science,\\ 2A $\&$ 2B Raja S.C. Mullick Road, Kolkata 700032, India}

\author{Sk Jeesun}
\email{skjeesun48@gmail.com}
\affiliation{School of Physical Sciences, Indian Association for the Cultivation of Science,\\ 2A $\&$ 2B Raja S.C. Mullick Road, Kolkata 700032, India}
\begin{abstract}
The basic idea of this work is to achieve the observed relic density of a non-thermal
dark matter(DM) and its connection with Cosmic Microwave Background (CMB) via 
additional relativistic degrees of freedom which are simultaneously generated during 
the period $T_{\rm BBN}~{\rm to}~T_{\rm CMB}$ from a long-lived dark sector particle.
To realize this phenomena we minimally extend the type-I seesaw scenario with a Dirac 
fermion singlet($\chi$) and a complex scalar singlet ($\varphi$) which transform 
non-trivially under an unbroken symmetry $\mathcal{Z}_3$. $\chi$ being the lightest 
particle in the dark sector acts as a stable dark matter candidate while the next to 
lightest state $\varphi$ operates like a long lived dark scalar particle. The initial 
density of $\varphi$ can be thermally produced through either self-interacting number 
changing processes ($3 \varphi \to 2 \varphi$) within dark sector or the standard 
annihilation to SM particles ($2 \varphi \to 2~ {\rm SM}$). 
The late time (after neutrino decoupling) non-thermal decay of $\varphi$ can produce 
dark matter in association with active neutrinos. The presence of extra
relativistic neutrino degrees of freedom at the time of CMB can have a 
significant impact on $\Delta \rm N_{eff}$. Thus the precise measurement of 
$\Delta \rm N_{ eff}$ by current PLANCK 2018 collaboration and future experiments 
like SPT-3G and CMB-S4 can indirectly probe this non-thermal dark matter scenario
which is otherwise completely secluded due to its tiny coupling with the standard 
model. 
\end{abstract}

\keywords{SIMP, WIMP, FIMP and CMB}
\maketitle
\section{Introduction}
The standard model (SM) of particle physics has been extraordinarily 
victorious in explaining properties of  elementary particles of the universe and their interactions
through strong, electromagnetic and weak forces. The SM seems complete after the 
discovery of Higgs-like particle with mass $M_h = 125 $ GeV at the Large Hadron Collider (LHC)\cite{CMS:2012qbp,ATLAS:2012yve}, which is 
responsible for mass generation mechanism through electroweak symmetry 
breaking in the SM. Inspite of the great triumph of the SM, 
several theoretical and experimental issues still persist, that demands
physics beyond the framework of the Standard Model. 
Based on numerous astrophysical and cosmological observations at a wide 
range of length scales, it is now well established  
fact that about 80$\%$ of total mass of the universe consists of 
Dark matter (DM)\cite{Zwicky:1933gu,Rubin:1970zza,Clowe:2006eq,Planck:2018vyg}
with relic density $(\Omega_{\rm DM}h^2=0.120\pm 0.001)$~\cite{Planck:2018vyg}.
Another astonishing experimental evidence is the observation of tiny but non-zero neutrino
masses $(m_\nu \lesssim {\mathcal O}(10^{-10})~{\rm GeV})$ and neutrino flavour oscillations 
\cite{T2K:2011ypd, DoubleChooz:2011ymz,DayaBay:2012fng,RENO:2012mkc,
MINOS:2013xrl, ParticleDataGroup:2020ssz}. To address these issues,
various theoretical as well as phenomenological ideas have been proposed.
The issue of neutrino masses and their mixing angles can be resolved by the
Seesaw mechanisms \cite{Minkowski:1977sc,Mohapatra:1979ia,
Schechter:1980gr,Gell-Mann:1979vob, Mohapatra:1980yp,Lazarides:1980nt,
Wetterich:1981bx,Schechter:1981cv,Brahmachari:1997cq,Foot:1988aq}. However,
any direct experimental verification of these ideas are yet to be confirmed.
While in the dark matter sector, weakly interacting massive particles 
(WIMP) \cite{Kolb:1990vq,Feng:2010gw,Roszkowski:2017nbc,Schumann:2019eaa,
Lin:2019uvt,Arcadi:2017kky} is the most popular and widely studied thermal 
DM candidate whose interaction strength with SM particles is of the order 
of electroweak interactions and via freeze-out mechanism it fits nicely the observed relic density of 
the universe. Nevertheless, null 
measurements from various dark matter detection experiments \cite{PandaX-II:2016vec, XENON:2017vdw, LUX:2016ggv,PICO:2019vsc,
AMS:2013fma,Buckley:2013bha,Gaskins:2016cha,Fermi-LAT:2016uux,MAGIC:2016xys,
Bringmann:2012ez,Cirelli:2015gux,Kahlhoefer:2017dnp,Boveia:2018yeb} severely 
restricts the WIMP freeze-out mechanism and forcing us to think if the 
standard WIMP paradigm is just waning or it is already deceased. 
To bypass this deadlock, an alternative framework, coined as 
{\it freeze-in} mechanism has been proposed. In this framework DM is a
feebly interacting massive particle (FIMP) whose interactions with SM 
plasma is too small $\lesssim$ $\mathcal{O} (10^{-10})$ 
to keep them in thermal bath \cite{Hall:2009bx,Konig:2016dzg,
Biswas:2016bfo,Biswas:2016iyh, Bernal:2017kxu,Borah:2018gjk}.
Rather FIMPs are produced non-thermally either from decay or annihilation 
of bath particles in the early universe. The FIMP freezes in once the
temperature of the universe becomes lower than the FIMP mass 
and produces DM relic abundance in the correct ball-park as 
observed today. Moreover, FIMPs having such a petite coupling 
with SM particles can easily accommodate various non-observational 
signature of DM in different detection experiments 
like Panda \cite{PandaX-II:2016vec}, 
XENON \cite{XENON:2017vdw}, LUX \cite{LUX:2016ggv}. However, some 
attempts have been made to test the FIMP scenario in direct search experiments \cite{Elor:2021swj,Bhattiprolu:2022sdd}, and indirectly using 
observational data from big bang nucleosynthesis (BBN) or cosmic microwave 
background (CMB)~\cite{Heeck:2017xbu,Bae:2017dpt,Boulebnane:2017fxw,
Ballesteros:2020adh,DEramo:2020gpr,Decant:2021mhj,Li:2021okx}.
Furthermore, non-thermal production of DM from the 
decay of heavier dark sector particles have also been studied in 
literature \cite{Cheng:2020gut,Chu:2011be, Ghosh:2022fws,Hambye:2019dwd}.  

Apart from FIMP, strongly interacting massive particle(SIMP) is another alternate paradigm to explain the DM abundance 
\cite{Hochberg:2014dra,Hochberg:2014kqa,Bhattacharya:2019mmy} as well as 
the structure 
formation of the universe\cite{Randall:2008ppe, deBlok:2009sp, Boylan-Kolchin:2011qkt}. 
SIMPs are produced thermally in the early universe by number changing processes within itself.
SIMP scenario requires strong self interaction and very small annihilation rate to SM particles contrary to WIMPs to successfully satisfy the
correct DM relic density \cite{Bernal:2018ins,Bernal:2015ova,Ghosh:2021wrk}.

On the other hand Cosmic Microwave Background (CMB) is an ideal probe of the physics in the
early universe. The very precise measurement of anisotropies in 
the temperature of photons which dissociate from visible sector in the 
recombination phase of the thermal evolution of our universe, leads to 
the determination of the energy density in that particular 
era. From this 
one can estimate the number of light species in the universe and in the 
massless limit this is provided by the relativistic degrees of freedom
$g_{*}$ \cite {Khuri:1998xx,Riemer-Sorensen:2013iql}. On the other hand after neutrino decoupling,
one recasts the number of light degrees of freedom associated with
neutrino bath as ${\rm N_{eff}}$ and in the SM it is roughly number of active 
neutrinos $(N_\nu = 3)$.  Thus any physics scenarios beyond the SM (BSM) with new light degrees of
freedom with masses ${\cal O}~(\rm eV)$ or less can subscribe to 
$\rm N_{eff}$. We have very precise information of $\rm N_{eff}$ 
from recent Planck 
2018 \cite{Planck:2018vyg}, which suggests $\rm N_{ eff}$ at the time of 
CMB formation to be ${\rm N_{eff}^{CMB}=2.99^{+0.34}_{-0.33}}$ at 
95$\%$ confidence level (C.L), whereas in the SM, 
$\rm N^{ SM}_{\rm eff} = 3.045 $.
The quantity $\rm N_{eff}$ is parameterized as ${\rm N_{eff}}\equiv {\left(\rho_{rad}-\rho_{\gamma}\right)} /{\rho_{\nu}}$, where,
$\rho_\gamma $, $\rho_\nu $, and  $\rho_{\rm rad} $ denote the photon energy density, active neutrino energy density and total radiation energy density of the universe 
respectively \cite{Mangano:2005cc}.
The deviation from 3, the number of active neutrinos can be attributed to various non-trivial effects like non-instantaneous neutrino decoupling, 
finite temperature QED corrections to the electromagnetic plasma, flavour oscillations of neutrinos \cite{Mangano:2005cc,Grohs:2015tfy,deSalas:2016ztq,Akita:2020szl}. 
Multiple upcoming
experiments like SPT-3G\cite{SPT-3G:2019sok}, 
CMB-S4\cite{CMB-S4:2016ple} are going to be extremely sensitive to the presence of any 
new radiation /light degrees of freedom
and will put 
stringent bound on $\rm{\Delta N_{eff}}=  N_{\rm eff}-N_{\rm eff}^{\rm SM}\approx 0.06$ at 95\% confidence level. 
There also exists strong constraint on the changes in $N_{\rm eff}$ between BBN and CMB epoch as explored by ref.\cite{Yeh:2022heq}.
Various BSM scenarios that entail additional entropy injection to the neutrino sector can face a tough challange from the measurement of $\Delta {\rm N_{\rm eff}}$ by
both the present and future generation CMB experiments \cite{Boehm:2012gr,Brust:2013ova,Nollett:2014lwa,Abazajian:2019oqj,Borah:2022obi,Biswas:2022fga}.
This precise measurement of $\Delta {\rm N_{\rm eff}}$ has also non-trivial implications on various new physics models that produce dark matter in associated with the injection of 
additional light degrees of freedom \cite{Knapen:2017xzo,Choudhury:2019sxt,Biswas:2022vkq,Ganguly:2022ujt}. Recently, studies with $\Delta {\rm N_{\rm eff}}$ have been explored 
in the context of resolving the discrepancy between local measurement and CMB estimation of the Hubble constant ($H_0$) \cite{Vagnozzi:2019ezj,deJesus:2022pux}.

In this work, we are interested in non thermal production of dark matter from heavier dark sector, where the dark sector may or may not have sizeable interaction with the SM bath.
To realize this picture we extend the SM by  one complex SM gauge singlet scalar ($\varphi$), one gauge singlet Dirac fermion ($\chi$) and 3 right handed neutrinos (RHN)($N_{1,2,3}$). 
The three RHNs are responsible for neutrino mass generation through well known Type-I seesaw mechanism\cite{Mohapatra:1979ia,Ibarra:2010xw}. 
$\varphi$ and $\chi$ are dark sector particles and an additional discrete $\mathcal{Z}_3$ symmetry has been imposed under which they transform non trivially while the rest of the particles transform 
trivially.
In our analysis lightest dark sector particle $\chi$ can play the role of DM whereas the heavy dark sector particle ($\varphi$) is a long lived owing to 
its very small coupling $(\lesssim 10^{-12})$, which will eventually allows
$\varphi \to \chi \nu $ decay at temperature below neutrino decoupling 
temperature ($\sim1~{\rm MeV}$).
Non thermal decay of $\varphi$ is the only source of DM($\chi$) production 
whereas $\varphi$ freezes out thermally and gains non-zero 
number density via either of these two mechanisms : $(i)$ the number changing 
self interactions ($3\varphi \to 2\varphi$), $(ii)$ annihilation to SM particles 
($2\varphi \to 2{\rm SM}$). In this work we emphasise on the first scenario where $\varphi$ has strong self 
interactions but very weak interaction with the SM bath. The implication of
this particular scenario has been so far overlooked. Through our detailed 
numerical analysis we will highlight the importance of this mechanism in 
both DM phenomenology and its footprint on CMB. For the shake of completeness of the analysis, 
we will also consider the second process as well to showcase region of parameter space where 
these two scenarios are relevant. 

It should be noted that in both cases $\varphi$ particle maintain 
kinetic equilibrium with SM bath via the elastic scattering processes
and share common temperature with SM bath contrary to studies that deal with secluded or decoupled dark sector scenarios \cite{Farina:2016llk,Pappadopulo:2016pkp}.
If the decay of $\varphi$ is happening after neutrino decoupling then it will increase neutrino bath entropy and contribute to $\rm \Delta N_{eff}$.
If the decay is completed before CMB we can trace the signature of DM from $\rm \Delta N_{eff}$ at the time of CMB and find some interesting correlation of freeze in DM  and  $\rm \Delta N_{eff}$ in our proposed set up. 
{Analogous scenarios of entropy injection between BBN and CMB epoch have been already investigated in various contexts. These include 
the consequence on priomodial nucleosynthesis resulting from the 
decay of non relativistic particle \cite{Scherrer:1987rr}, the presence of 
relativistic dark matter from long lived state\cite{Hooper:2011aj}, and 
the contribution of dark radiation arising from decyaing particle
\cite{Fischler:2010xz,Menestrina:2011mz,Hasenkamp:2012ii,Sobotka:2022vrr}.
Although the aforementioned papers have conducted extensive analysis, most
of them rely on a model independent formalism. In contrast, our analysis 
focuses on a specific particle physics model that aims to elucidate 
the observed DM relic density along with the production mechanism of DM. Among the examples
listed above, only reference \cite{Hooper:2011aj} discusses nonthermal DM,
which can account for a small fraction ($\sim 1\%$) of the 
observed relic density. Conversely, our proposed framework provides nonthermal 
DM from the decay of self interacting heavier particle within the 
dark sector, providing a comprehensive explanation for the entire observed 
relic density.}

The rest of the paper is structured as follows: In section \ref{sec:model} we introduce the model. The possible dynamics of DM production have been discussed
in section \ref{sec:dynamics}. In section  \ref{sec:neff} we discuss the light neutrino production from late time decay of $\varphi$. The outcome of DM relic density 
together with the contribution to $\rm \Delta N_{eff}$ at CMB for both scenarios-I and II  have been discussed in section \ref{sec:relic_neff}. Finally, 
we summarize our results in section \ref{sec:conc}. We show relevant theoretical constraints and limit from the SM Higgs invisible decay width 
in Appendix \ref{sec:tc} and Appendix \ref{sec:higgs} respectively. Feynman diagrams and corresponding thermal averaged cross-section for
$3\varphi \to 2\varphi$ and $2\varphi \to 2 {\rm SM}$ processes are explicitly demonstrated in Appendix \ref{sec:3DMto2DM} and  Appendix \ref{sec:2DMto2SM} respectively. A brief discussion on kinetic equilibrium is mentioned in Appendix \ref{sec:KEQ}.

\section{The Model}
\label{sec:model}
In order to explain DM production from  dark sector and its cosmological imprints in CMB, we extend the SM by a  complex scalar $\varphi$, one Dirac fermion $\chi$ and three neutral Majorana fermions, $N_{1,2,3}$ which are singlet under the SM gauge group.
An additional $\mathcal{Z}_3$ symmetry provides the stability of the lightest dark sector particle, under which the field $\varphi$ and  $\chi$ transform non-trivially 
 i.e. $\{\varphi,~\chi \} \to \{ e^{i\frac{2\pi}{3}} \varphi,~e^{i\frac{2\pi}{3}} \chi\}$ while all the SM fields including $N_{1,2,3}$ transform trivially 
i.e. $\{N_{1,2,3},~{\rm SM} \} \to \{ N_{1,2,3},~{\rm SM} \} $
\footnote{In general any $\mathcal{Z}_N$ symmetry can serve similar kind of scenario with different self interacting number changing processes, $m~\varphi \to 2~\varphi$ ($m \geq 3$), as well as the standard annihilation to SM particles, $2\varphi \to 2 {\rm SM}$. 
For example, $\mathcal{Z}_2$ will provide $4 \varphi \to 2 \varphi $ interactions which are more phase space suppressed for $M_\varphi \sim \mathcal{O}({\rm MeV})$ compare to $3 \varphi \to 2 \varphi$ interactions realised in $\mathcal{Z}_3$ symmetry \cite{Hochberg:2014dra}.}. 
 The lightest dark state, $\chi$ acts as a stable DM candidate which is produced from the late time decay of the other dark sector particle, $\varphi$. The right handed neutrinos (RHN) i.e. $N_{1,2,3}$ which do not transform under $\mathcal{Z}_3$,  will be responsible for light neutrino mass via Type-I seesaw mechanism \cite{Mohapatra:1979ia}.
All the BSM fields and their corresponding charge assignments under the extended SM Electroweak (EW) gauge group are tabulated in table-\ref{tab_1}.

\begin{table}[htb!]
\begin{tabular}{ |c|c|c|c|c| } 
 \hline
 \hline
 \multicolumn{2}{|c}{BSM Fields} &  \multicolumn{1}{|c|}{$SU(2)_L$} & $U(1)_Y$ & $\mathcal{Z}_3$\\
   \hline
   \hline
  \makecell{Dark scalar (DS)}& $\varphi$ & 1 & $0$ & $\omega (\equiv e^{i \frac{2 \pi}{3}})$ \\ 
  \hline
  DM & $\chi$ & 1 & $0$ & $\omega (\equiv e^{i \frac{2 \pi}{3}})$ \\ 
  \hline
  \hline
  RHN& $N_{1,2,3}$ & 1 & $0$ & $1$ \\
  \hline
  \hline
 \end{tabular}
\caption{\it Charge assignment of BSM fields under the extended SM EW gauge group, $\mathcal{G}_{\rm SM}^{\rm EW} \otimes \mathcal{Z}_3$.}
\label{tab_1}
\end{table}
The Lagrangian of this model takes the following form :
\begin{eqnarray}
 \mathcal{L}&=& \underbrace{\mathcal{L}^{\rm K+Y}_{\rm SM}-V(H)}_{\rm SM}+
 {\mathcal{L}_{\rm N}+\mathcal{L}_{\rm DS}+\mathcal{L}_{\rm DS-H}+\mathcal{L}_{\rm DS-\nu}} ~.
\end{eqnarray}
Here, $V(H)$ represents the SM Higgs potential which is given by
\begin{eqnarray}
V(H)&=&-\mu_H^2 |H|^2 + \lambda_H |H|^4 . 
\end{eqnarray}
The BSM part encapsulate interactions of heavy RHN sector
($\mathcal{L_{\rm N}}$), dark sector($\mathcal{L}_{\rm DS}$) as well as 
their connection with the SM.  
The interaction of heavy RHN sector is described by,
\begin{eqnarray}
\mathcal{L}_{\rm N}=&& \sum_{i}
i \Bar{N_i} \gamma^{\mu}\partial_{\mu}N_i - \sum_{i,j}\frac{1}{2}M_{N_{i j}} \Bar{N_i^c}N_{j}
-\sum_{\ell, j}Y_{\ell j} \Bar{L_{\ell}} \Tilde{H} N_{j}
+h.c.  
\label{eq:RHN}
\end{eqnarray}
where $i,j=1,2,3$ and $\ell=e,\mu,\tau$ are lepton flavour indices. 
$L_\ell=(\nu_\ell~~ \ell )^T$ are left handed the SM lepton doublet and 
$H$ is the SM scalar doublet with $\Tilde{H}=i\sigma_2 H^*$.
The second term in eq.\eqref{eq:RHN} is the Majorana mass term associated with $N_{1,2,3}$ and the last term is the Dirac Yukawa interactions with $N_{1,2,3}$.
After electroweak symmetry breaking (EWSB), the
SM scalar doublet, $H$ can be expressed in unitary gauge  as
$H=\left(\begin{matrix} 0 && \frac{h+v}{\sqrt{2}} \end{matrix}\right)^T$ where $v=246$ GeV is vaccum expectation value (VEV) of SM Higgs.
Active neutrino masses can be generated via Type-I seesaw mechanism followed from eq.\eqref{eq:RHN} as $
 \big(m_\nu \big)_{3\times 3} \approx \big({Y v}/{\sqrt{2}}\big) \big(M_N\big)^{-1} \big({Y^T v}/{\sqrt{2}}\big)
$ 
and the mixing angle between active neutrino and RHN is then
$
 \theta_{\rm mix} \sim \big(Y v/\sqrt{2}\big) \big( M_N \big)^{-1}~,
$ where $M_N \approx \big(M_N \big)_{3\times3} $ \cite{Mohapatra:1979ia}.

The dark sector of this model consists of a complex scalar ($\varphi$) and a Dirac fermion ($\chi$) with similar transformation property under $\mathcal{Z}_3$. The lightest state behaves as a stable DM particle. The Lagrangian of the dark sector is described as follows:
\begin{eqnarray}
    \mathcal{L}_{\rm BSM}&\supset& \mathcal{L}_{\rm DS}+\mathcal{L}_{\rm DS-H}+\mathcal{L}_{\rm DS-\nu} \nonumber \\
    &=& \Big( |\partial_{\mu}\varphi|^2 -\mu^2 |\varphi|^2  +  i \Bar{\chi} \gamma^{\mu}\partial_{\mu}\chi - M_{\rm DM} \Bar{\chi} \chi
 -{\lambda_{\varphi}} |\varphi|^4
-\frac{\mu_{\varphi}}{3!}(\varphi^3 +\varphi^{*3})   -y_{_{\varphi\chi}} \overline{\chi^c}\chi \varphi  \Big) \nonumber \\
&& + \Big(- \lambda_{\varphi H} |H|^2|\varphi|^2 \Big)  + \Big(-\sum_i y_{_{\varphi N_i}} \Bar{\chi} \varphi N_{i} + h.c. \Big)~~,
\label{eq:lagDS}
\end{eqnarray}
 where, $i=1,2,3$. In the above equation, $\mu$ is the bare mass term of  $\varphi$ and $M_{\rm DM}$ is the mass of dark fermion  $\chi$. 
 For simplicity, in this work we consider all parameters to be real. 
 In the dark scalar sector, we assume $\mu >0$ and $\lambda_\varphi > 0$ so that $\langle \varphi \rangle =0 $  which implies unbroken $\mathcal{Z}_3$ symmetry. After EWSB the physical mass of $\varphi$ can be expressed as,
\begin{equation}
M_{\varphi}^2= \mu^2 + \frac{\lambda_{\varphi H}\,v^2}{2}   ~.
\end{equation}
The most important interaction as far as our analysis is concerned, is
given by the Yukawa interaction involving the dark scalar $(\varphi)$,
the DM $(\chi )$ and SM neutrinos $(\nu)$:
\begin{align}
 \mathcal{L}^{\rm int}_{\rm DS-\nu}= y_{_1}\overline{\chi} \nu \varphi  + h.c.
 \label{eq:yukdm}
 \end{align}
This Lagrangian can be realized from the last term in braces in eq.\eqref{eq:lagDS} via small mixing angles($\theta_{\rm mix}$) with RHNs($N_{1,2,3}$). The effective Yukawa coupling, $y_1$ can be understood as $\sum_{i} y_{_{\varphi N_i}} \theta_{\rm mix}^i$, where $i=1,2,3$. 
 
\begin{figure}[h]
\centering
\includegraphics[scale=0.45]{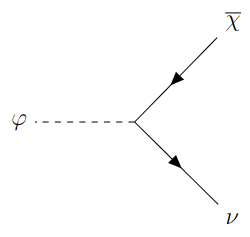} 
 \caption{\it Diagram of DM production with active neutrinos from NLP $\varphi$}
\label{fig:c1}
\end{figure}

We choose our model parameters of the dark sector, in such a manner 
that we always get 
$\chi$ as the lightest dark sector particle. This mass pattern and 
the underlying discrete symmetry ensure us that
the Dirac fermion $(\chi)$ with mass $M_{\rm DM}$ is the DM particle
and $\varphi$ with mass $M_\varphi$ is the next to 
ligtest particle (NLP) in this framework. The DM interacts with the SM 
bath only through $\varphi $ via the Yukawa interaction shown in eq.(\ref{eq:yukdm}). 
Thus for a given mass hierarchy between $\varphi$ and 
$\chi$, the life-time of $\varphi$ is determined by the strength of the Yukawa coupling $y_{_1}$. 
For our analysis, we assume NLP $(\varphi)$ to be a long-lived $(\tau_\varphi > \tau_{\rm BBN})$ 
particle and for this to happen one requires a very tiny Yukawa coupling 
$y_1 \lesssim 10^{-12}$ (for $M_{\varphi}\sim \mathcal{O}$(GeV)).
The NLP $\varphi$ can be thermally produced via the sizable Higgs-portal interaction or through number changing self interaction processes.
The production of the DM in the thermal bath through scattering process
is highly suppressed because of it feeble coupling ($y_1$). However, it can be produced non-thermally from the decay of long-lived $\varphi$ as shown in Fig.\ref{fig:c1}.
\noindent The decay width of $\varphi$ to DM and a light neutrino is given by, 
\begin{eqnarray}
\Gamma_{\varphi\rightarrow \overline{\chi}\nu}=\dfrac{y_1^2 \,M_{\varphi} }{16\, \pi}  \left(1 - \dfrac{ M_{\rm DM}^2}{ M_{\varphi}^2} \right)^2.
\label{eq:decay}
\end{eqnarray} 
Besides this, there are two more production channels of the DM $\chi$:
$(a)$ $N_{1,2,3} \to \chi  \varphi$ and $(b)$
$\varphi \to {\bar \chi^c} \chi$.
The main aim of this work is to connect non-thermal DM and $\Delta N_{\rm eff}$ producing from self-interacting dark sector (NLP) which is achievable via the decay $\varphi \to \chi \nu$. 
But the presence of those new channels $(a~\& b)$ 
will dilute the effect of the late time decay of $\varphi$ in $\Delta N_{\rm eff}$
and may even completely imperil our non-thermal dark matter 
scenario by thermalizing the dark sector.  
To avoid DM production from RHNs we set $M_{N_{1,2,3}} \gg T_{\rm RH}$
so that their number densities get Boltzmann suppressed($e^{-M_N/T}$)
\cite{Ganguly:2022ujt}. 
Therefore for our discussion, we choose the following hierarchy 
\begin{equation}
M_{N_{1,2,3}} \gg T_{\rm RH} >  M_{\varphi} > M_{\rm DM} .
\label{eq:t_rh}
\end{equation}
In order to get active neutrino mass of the order $\sim 0.1 $ eV, we require 
$M_{N_{1,2,3}}\sim\mathcal{O}(10^{10})$ GeV and 
$\theta_{\rm mix}\sim \mathcal{O}(10^{-10})$ \cite{Kersten:2007vk} and
to satisfy the criteria of eq.\ref{eq:t_rh} we set $T_{\rm RH} =10^3 $ GeV
which is much above the current bound obtained from BBN \cite{Hannestad:2004px}
\footnote{{The current lower bound on $T_{\rm RH}$ from BBN is few MeV \cite{Hannestad:2004px} and succesfull baryogenesis is possible even with $T_{\rm RH} \lesssim 100$ GeV \cite{Davidson:2000dw,Jaeckel:2022osh}. However, in that case one also has to set an upperbound on $M_{\varphi} \ll T_{\rm RH}$ in context of our analysis to get sufficient $\varphi$ number density. On the other hand the effect of $T_{\rm RH}$ on primordial gravitational waves depends on specific inflationary model \cite{Nakayama:2008wy,Figueroa:2019paj,Artymowski:2017pua} whose analysis is beyond the scope of this work.}}.
Following this argument and masses of relevant particles of this model, 
in the rest of our analysis we can safely ignore the production of DM from 
RHN decay in the computation of $Y_\varphi$ and $Y_\chi$. Moreover 
to suppress the process $(b)$ we consider $y_{_{\varphi\chi}} \ll y_{_1}$, 
and this is necessary to exalt $\varphi \to \chi \nu$ decay 
so that $\varphi$ can have the maximal contribution to $\rm \Delta N_{eff}$.

\noindent Interestingly, active neutrinos($\nu$) produced from the decay of NLP $\varphi$ 
along with DM($\chi$) as shown in Fig.\ref{fig:c1} can have very 
intriguing consequences in the observation of CMB. 
We assume the value of Yukawa coupling ($y_{_1}$) such that $\varphi \to \chi \nu$
decay mostly happens between neutrino decoupling temperature (${\rm T} <2~{\rm MeV}$) 
and CMB formation  (${\rm T}\approx {\rm 0.1 eV}$). This promptly opens
up the possibility of probing the impact of extra neutrino production from CMB 
radiation. And this can be achieved if $y_{_1}$ varies
in the range $(10^{-12}-10^{-15})$ and for such a tiny coupling
$\varphi$ becomes a long-lived particle ($\tau_{\varphi}>\tau_{\rm BBN}$).
Thus the aforementioned supplementary active neutrino ($\nu$) injection in our proposed 
scenario increases neutrino sector entropy and which in turn 
contribute significantly to additional neutrino degrees of freedom or 
$\Delta \rm N_{eff}$ which is very precisely measured at the time of CMB.
Thus any experimental observation on $\Delta \rm N_{eff}$ can have 
very intriguing impact on the dynamics of dark scalar $\varphi$ which 
in turn can influence the dark matter $(\chi)$ abundance via 
$\varphi \to \chi \nu$ decay process, thus affecting two 
disjoint (FIMP dark matter \& $N_{\rm eff}$) sectors simultaneously. 
To explore this phenomenology, we 
perform a detailed numerical scan over model parameters 
to show that the precise 
measurement of $\rm \Delta N_{eff}$ at CMB can indeed restrict certain 
region of parameter space of non-thermal DM production which is 
otherwise remains elusive to visible sector due to 
extremely tiny 
strength of interactions involved in such non-thermal DM production process.

\noindent While doing our numerical analysis, we use the following model parameters: 
\begin{equation}
 \{  M_{\rm DM},~M_{\varphi},~\lambda_{\varphi H},~\lambda_{\varphi},~\mu_{\varphi},~ y_{_1}   \},
\end{equation}
Here, the Higgs portal coupling $\lambda_{\varphi H}$
which decides the interaction between $\varphi$
and SM, plays a 
significant role in deciding $\varphi$'s number density through 
$2\varphi\to  2~{\rm SM}$ annihilation and also in
$(\varphi~{\rm SM} \to \varphi~{\rm SM})$  elastic scattering processes.
On the other hand, the scalar sector parameters $\lambda_{\varphi}$ and $\mu_{\varphi}$ decide the self 
interactions of $\varphi$ which is relevant for the number 
changing processes like $3\varphi \to 2 \varphi $.
And finally the effective Yukawa coupling,
$y_1$ 
dictataes both DM abundance and additional contribution to
$ N_{\rm eff}$. 
%
\section{Dynamics of dark sector}
\label{sec:dynamics}
In this section, we discuss the dynamics of the dark sector 
that leads to the early time production of the heavy NLP dark 
scalar $(\varphi)$ followed by the late time non-thermal production  
of DM $(\chi)$ from the decay of $\varphi$. The number density of
DM will be generated at some later epoch (after the neutrino
decoupling temperature) of the Universe via the 
following two steps :

\begin{itemize}
 \item {Step I}:~~ {\it thermal production } of heavy dark scalar ($\varphi$) at 
the early time of Universe ($t <\tau_{\rm BBN}$).
 \item {Step II}:~~{\it non-thermal production} of DM ($\chi$) from the 
late time decay of $\varphi$ ($\tau_{\rm BBN}< t < \tau_{\rm CMB}$).
\end{itemize}

\begin{figure}[h]
\centering
\includegraphics[scale=0.32]{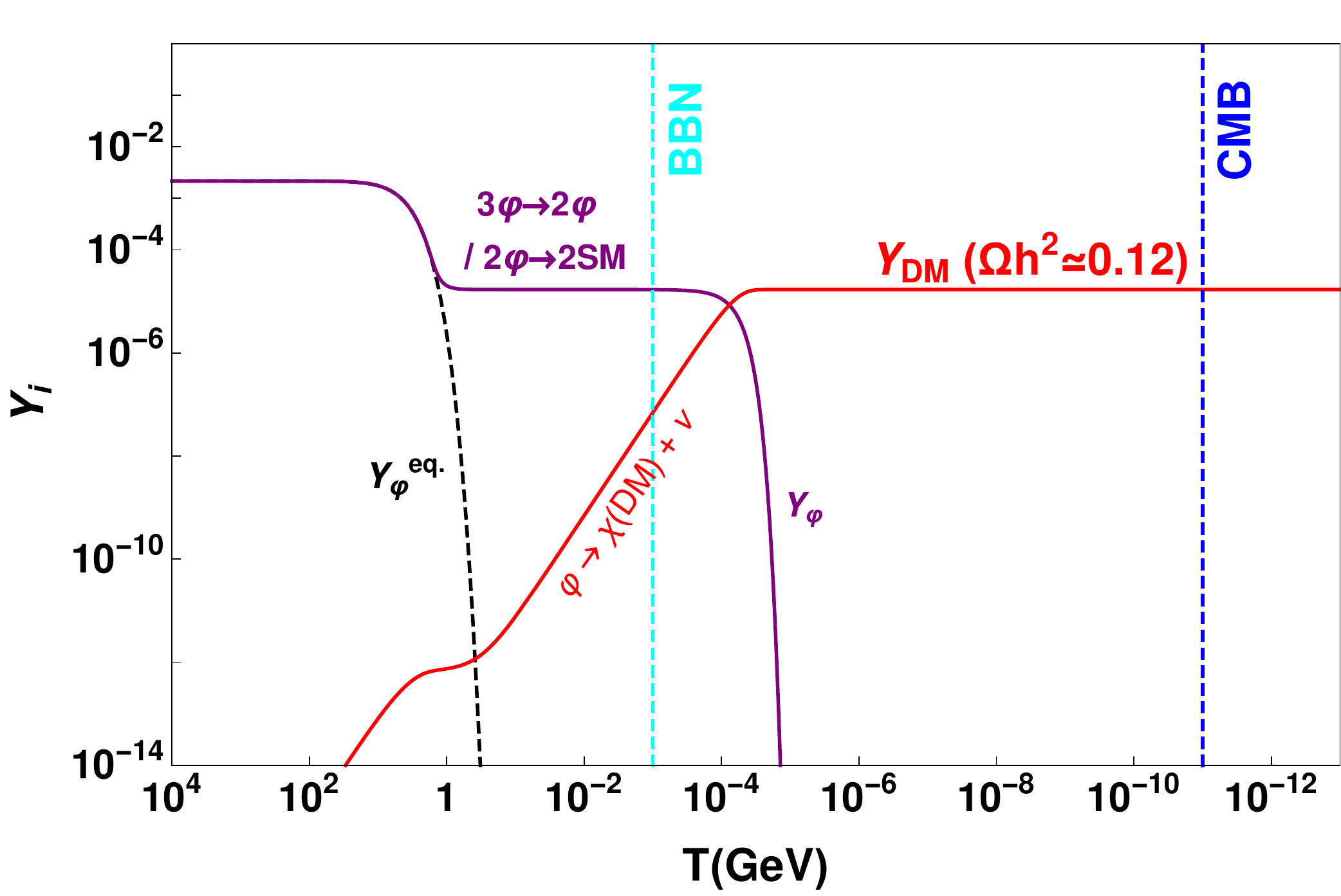} ~~
\includegraphics[scale=0.22]{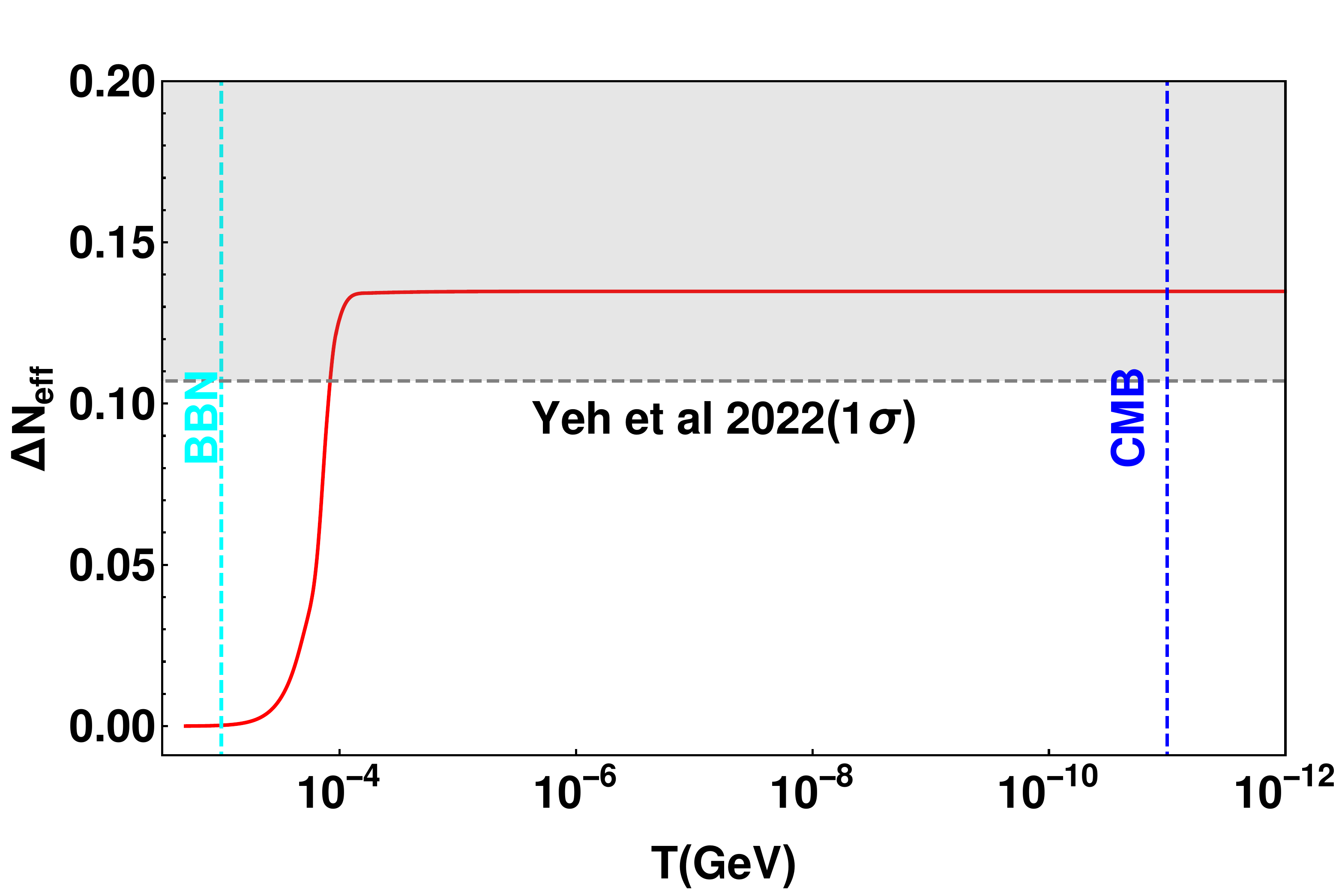}
 \caption{\it A cartoon diagram of DM production(left) and the impact in $\rm \Delta N_{eff}$ at the time of CMB(right).}
\label{fig:c2}
\end{figure}
A cartoon diagram of our proposed setup is shown in Fig.\ref{fig:c2}.
In the left panel, we show the variation of co-moving density 
as a function of temperature. The purple and red solid lines correspond to 
the thermal production of $\varphi$ (Step I) and the 
non thermal production of DM $(\chi)$ (Step II)
respectively. We also show two important
temperatures, namely, the BBN and CMB that play crucial role in our analysis.
Active neutrinos produced in the aforementioned decay of $\varphi$ 
make substantial contributions to ${\rm N_{\rm eff}}$, which can attract 
severe constraints from various observational limits on $\rm \Delta N_{eff}$,
as shown in the right panel of Fig.\ref{fig:c2}. The gray rectangular band 
is excluded by the limit obtained in Yeh et al at $1\sigma $~\cite{Yeh:2022heq}. 
Having this broad picture in mind we now provide details of the thermal 
production of NLP followed by non-thermal production of DM in the rest of this
section.\\

\begin{figure}
\centering
\subfigure[\label{S1}]{\includegraphics[scale=0.50]{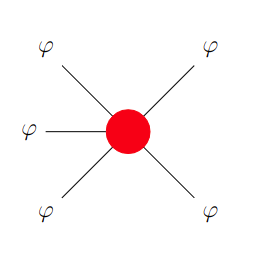}}~~~~~~~~
\subfigure[\label{S2}]{\includegraphics[scale=0.48]{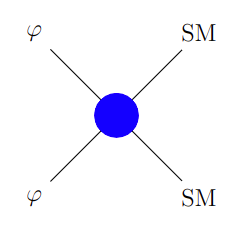}}
 \caption{\it A cartoon of number changing process of $\varphi$: $(a)$
three $\varphi$ annihilate to two $\varphi$ ($3\varphi \to 2\varphi$).
and $(b)$ two $\varphi$ annihilate to two SM particles ($2\varphi \to 2 {\rm SM}$).}
 \label{fig:cartoon1}
\end{figure}

{\bf{Step-I: Thermal production of $\varphi$}}\\
We consider a scenario in the early universe, when the interaction 
rate $(\Gamma^{\rm int}_\varphi)$ of the NLP $(\varphi)$ dominates over the 
expansion rate $(\mathcal {H})$ of the Universe, $(\Gamma^{\rm int}_\varphi >> 
\mathcal{H})$ so that $\varphi$ remains in thermal and chemical equilibrium.
As the temperature of the universe cools down, the interaction rate of 
$\varphi $ falls below the expansion rate of universe 
$( \Gamma^{int}_\varphi < \mathcal{H})$, thus the system departs from thermal equilibrium
and the number density of $\varphi$ freezes out. 
The number density of $\varphi$ is mainly provided by 
the following two types of number changing processes: 
$(i)$ ${3 \varphi \leftrightarrow 2 \varphi}$  via $\varphi$ self interactions
(shown in Fig.\ref{S1}) and 
$(ii)$ ${2 \varphi \leftrightarrow 2 ~{\rm SM} }$  via
the SM Higgs portal interactions (shown in Fig.\ref{S2}). 
As a result of these two number changing processes, the NLP $(\varphi)$ keeps its chemical 
equilibrium. On the other hand, the kinetic equilibrium is maintained 
between $\varphi$ and the SM bath via elastic scatterings, generically 
expressed as $\varphi ~ {\rm SM} \leftrightarrow \varphi ~ {\rm SM}$ 
which help $\varphi$ to keep same temperature with SM bath till freeze out
takes place. 

\noindent The complete dynamics of thermal production of $\varphi$ can be described by the
following Boltzmann equation: 
 \begin{eqnarray}
 \nonumber
 \dfrac{dY_{\varphi}}{dx}&=&- 0.116 \frac{g_{s}^2}{\sqrt{g_{\rho}}} \dfrac{M_{\varphi}^4}{x^5} M_{pl} \left< \sigma v^2\right>_{3\varphi \to 2 \varphi}(Y^3_{\varphi}-Y^2_{\varphi}Y^{eq}_{\varphi})\\
 \nonumber
 &&- 0.264 \frac{g_{s}}{\sqrt{g_{\rho}}} \dfrac{M_{\varphi}}{x^2} M_{pl} \left< \sigma v\right>_{2\varphi\rightarrow 2{\rm SM}}(Y^2_{\varphi}-{Y^{eq}_{\varphi}}^2)\\&&-\sqrt{\frac{45}{4 \pi^3}} \left< \Gamma _{\varphi \rightarrow \chi \nu} \right> \frac{x}{M_{\varphi}^2} \frac{M_{pl}}{\sqrt{g_{\rho}}}Y_{\varphi} ~.
 \label{eq:phi}
 \end{eqnarray}
Let us first describe various notations used in eq.\eqref{eq:phi}. $Y_\varphi(=\frac{n_\varphi}{s})$ is the co-moving number density of $\varphi$ where $s$ is the entropy density and $x$ is the dimensionless parameter defined as $x=\frac{M_\varphi}{T}$. ${ Y_{\varphi}^{eq}}$
is the equilibrium co-moving number density of $\varphi$.
$g_{s}(x)$ and $g_{\rho}(x)$ are the effective relativistic 
degrees of freedom associated with entropy density and the energy 
density respectively and finally $M_{pl}$ is the Planck 
mass($M_{pl}=1.22\times 10^{19}$GeV). The thermal averaged cross-section 
of $2\varphi \to 2~{\rm SM}$ process
is denoted by $\left< \sigma v\right>_{2 \varphi \rightarrow 2\rm{SM}}$ and 
for self-interacting number changing process $(3\varphi \to 2\varphi)$, it is 
defined as $\left< \sigma v^2\right>_{3\varphi \to 2\varphi}$. The first two terms 
in eq.\eqref{eq:phi} lead to non zero density of $\varphi$ via thermal 
freeze-out mechanism and it occurs at $x=x_F^{\rm tot.}$, 
where {\bf tot} in the superscript implies that both number 
changing processes i.e. ${3\varphi \to 2\varphi ~{\rm and}~ 2\varphi \to 2 {\rm SM}}$ 
are involved in $\varphi$ freeze-out process. 
The last term in eq.\eqref{eq:phi} provides the late time (after BBN) decay of $\varphi$  into DM $(\chi)$ and SM neutrinos,
resulting the dilution of number density of $\varphi$ into $\chi$ and $\nu$. 

\noindent From eq.\eqref{eq:phi} it is clear that 
two number changing processes of NLP $(\varphi)$ as discussed above 
are present to keep $\varphi$ in the thermal bath. However, depending upon 
the mass and couplings of NLP, it can be shown 
very easily that one of those two number changing 
processes is infact sufficient for the freeze-out and the final yield of 
NLP $(\varphi)$.
To justify our argument quantitatively  
we define the interaction rate of $3\varphi \to 2 \varphi$ process as:
$\Gamma_{3 \varphi \to 2 \varphi }=n_{\varphi}^2 \langle 
\sigma v^2\rangle_{3\varphi \rightarrow 2\varphi}$ and of $2\varphi 
\to 2 {\rm SM}$ as:  $\Gamma_{2 \varphi \to 2{\rm SM}}=n_{\varphi}\langle 
\sigma v\rangle_{2 \varphi \rightarrow 2\rm{SM}}$. In addition to these 
number changing processes, the number 
preserving scattering process ($\varphi ~{\rm SM} \to \varphi ~{\rm SM} $) is also 
present to keep $\varphi$ in kinetic equilibrium with SM bath. The interaction 
rate of this process is defined as: 
$\Gamma_{[\varphi~{\rm SM} \to \varphi~{\rm SM}]}=n_{\rm SM} \langle \sigma v 
\rangle _{[\varphi~{\rm SM} \to \varphi~{\rm SM}]}$. Depending on the relative 
interaction strength between two number changing processes of $\varphi$, we are interested in the  following two production modes of $\varphi$:
\begin{eqnarray}
 {\rm Scenario~I}&:&~~ \Gamma_{[\varphi~{\rm SM} \to \varphi~{\rm SM}]} ~>~ \Gamma_{3\varphi \to 2\varphi} ~\gg ~\Gamma_{2\varphi \to 2{\rm SM}}~~ , \nonumber \\
 {\rm Scenario~II}&:&~~ \Gamma_{[\varphi~{\rm SM} \to \varphi~{\rm SM}]} ~>~ \Gamma_{2\varphi \to 2{\rm SM}} \gg ~ \Gamma_{3\varphi \to 2\varphi} ~~ . \nonumber 
\end{eqnarray}

In the above hierarchy of scattering processes,  
 $\Gamma_{[\varphi~{\rm SM} \to \varphi~{\rm SM}]}$ plays a decisive part in 
maintaining kinetic equilibrium of $\varphi$. 
 During the freeze-out of $\varphi$ through processes 
like: $n \varphi \to 2 \varphi$, (for $n>2$) the rest mass energy of initial 
state particles can significantly enhance the kinetic energy of final
state particles, which in turn can heat up the dark sector 
\cite{Bernal:2015ova}, leading to an imbalance between the dark sector 
temperature $(T_\varphi)$ and SM bath temperature $(T)$. Thus, in general, 
to take into account this temperature imbalance one should consider a
 new parameter $(T_\varphi)$ in the evolution equation of 
NLP number density $(Y_\varphi)$ \cite{Ganguly:2022ujt}. However, in our study 
we can avoid this paradigm
by considering kinetic equilibrium between 
 $\varphi$ and SM bath, i.e.  by taking $T_\varphi = T $ at least upto 
the temperature at which $\varphi$ freezes out
from the thermal bath. And to achieve this, 
$\Gamma_{[\varphi~{\rm SM} \to \varphi~{\rm SM}]}$ must be 
larger than interaction rate of the other processes as well as the expansion 
rate $\mathcal{H}$ of the universe  (i.e $\Gamma_{[\varphi~{\rm SM} 
\to \varphi~{\rm SM}]}|_{x_F} \gtrsim \mathcal{H}({x_F}) $) \cite{Bernal:2018ins}. 
Most importantly this condition must 
be satisfied in both Scenario~I and II.
The relevant Feynman diagrams and thermal averaged cross-sections 
for $3\varphi \to 2\varphi$, $2\varphi \to 2{\rm SM}$ and 
$\varphi ~{\rm SM} \to \varphi ~{\rm SM}$ processes are shown in Appendices \ref{sec:3DMto2DM} and \ref{sec:2DMto2SM}. 

\begin{itemize}
 \item {\bf Scenario I}: In this scenario we consider the 
interaction rate of $3 \varphi \to 2 \varphi$ number changing 
process ($\Gamma_{3\varphi \to 2\varphi}$)
is significantly higher than $2 \varphi \to 2 {\rm SM} $ process ($\Gamma_{2\varphi \to 2~{\rm SM}}$).  
Thus $3 \varphi \to 2 \varphi $ process 
successfully keeps $\varphi$ in thermal bath for longer duration in 
comparison to the process ${2\varphi \to {2~ \rm  SM}}$.
Hence freeze-out of 
$\varphi$ is mainly governed by the $3 \varphi \to 2 \varphi$ process and it
occurs at $x=x_F^{3\varphi \to 2\varphi} \approx x_{F}^{\rm tot} > x_F^{2\varphi \to 2{\rm SM}}$. 
Here $x_F^{3\varphi \to 2\varphi}$ ($x_F^{2\varphi \to 2{\rm SM}}$) signifies the inverse freeze out temperature 
of $\varphi$ when only ${3\varphi \to 2\varphi}$ (${2\varphi \to 2{\rm SM}}$) is considered.

\begin{figure}[tbh]
     \subfigure[\label{b1}]{
    \includegraphics[scale=0.39]{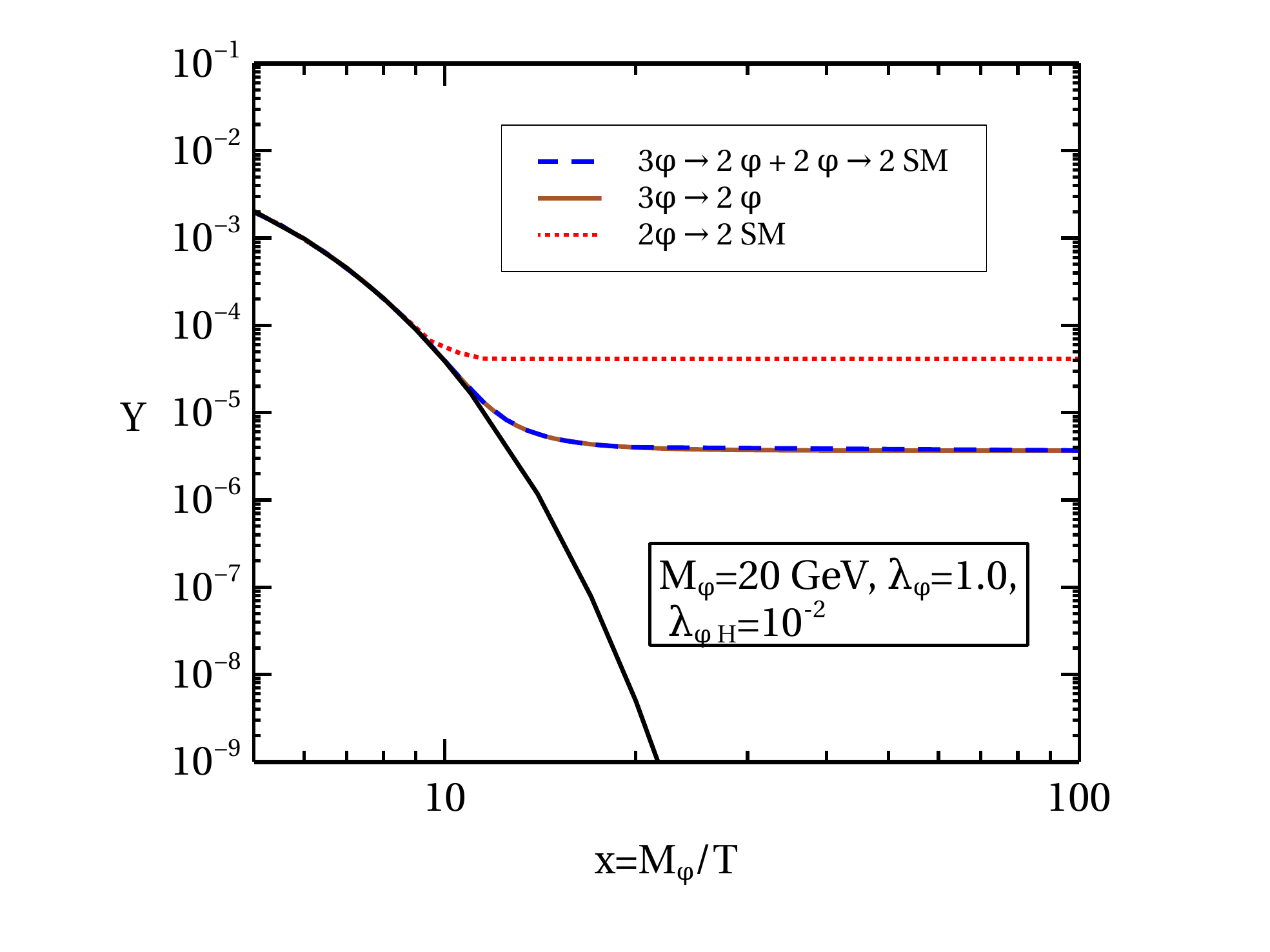}}
     \subfigure[\label{b2}]{
    \includegraphics[scale=0.39]{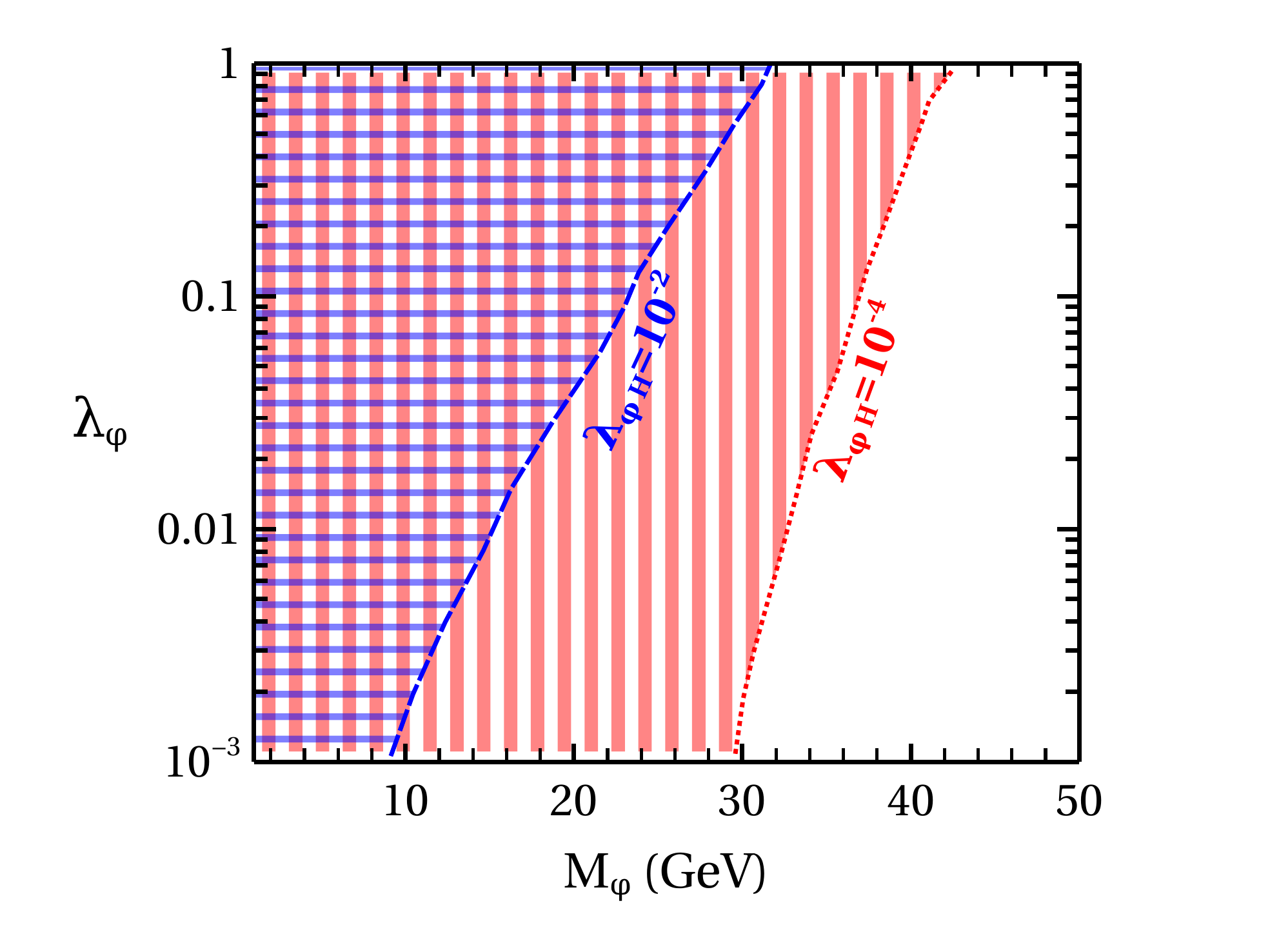}}
    \caption{{\it (a)Thermal freeze-out of $\varphi$ for $M_\varphi=20~ {\rm GeV},\mu_\varphi/M_\varphi=0.1,\lambda_{\varphi   H}=10^{-2},\lambda_\varphi=1.0$. The black solid line signifies $Y_{\varphi}^{eq}$.
    The blue dashed line corresponds to $Y_{\varphi}$ considering contributions from both $3 \varphi \to 2 \varphi$ and only $2\varphi \to 2~ {\rm  SM}$ processes. The brown solid line and red dotted line represent  $Y_{\varphi}$ considering only $3 \varphi \to 2 \varphi$ and $2\varphi \to 2~{\rm SM}$ process respectively.  (b)Parameter space for scenario-I in $M_{\varphi}$ vs.$\lambda_{\varphi}$ plane with $\mu_{\varphi}/M_{\varphi}=0.1$ for two different values of $\lambda_{\varphi H}=\{10^{-2},10^{-4}\}$ depicted by the blue and red shaded region respectively. The criteria for scenario-I  holds only for the region left to individual lines where $x_F^{\rm tot}$ is governed by  $3 \varphi \to 2 \varphi$  process.}}
    \label{fig:3to2}
\end{figure} 

In our model, the interaction rate of $3 \varphi \to 2 \varphi$ ($2 \varphi \to 2 {\rm SM}$) process  depends on the couplings 
$\lambda_{\varphi},\mu_{\varphi}/M_{\varphi}$ ($\lambda_{\varphi H}$ ) and  
mass $M_{\varphi}$.
To demonstrate the dynamics (where $\Gamma_{3 \varphi \to 2 \varphi} 
\gg \Gamma_{2\varphi \to {2~ \rm  SM}}$), we show the variation of 
the co-moving number density $Y_\varphi$ as a function of $x(=M_{\varphi}/T)$ 
in Fig.\ref{b1} for a sample Benchmark point:
$\{M_\varphi,~\mu_\varphi/M_\varphi,~\lambda_{\varphi H},~\lambda_\varphi\}=\{20~ {\rm GeV},~0.1,~10^{-2},~1\}$.
The black solid line corresponds to the equilibrium co-moving density 
 of $\varphi$ ($Y_{\varphi}^{eq}$) and the blue dashed line corresponds to 
the co-moving number density of $\varphi$ considering contributions from 
both the number changing processes: 
$3 \varphi \to 2 \varphi$ and $2\varphi \to 2~ {\rm  SM}$ in eq.\eqref{eq:phi}.
The brown solid line (red dotted line) depicts the variation of number 
density of $\varphi$ when only $3 \varphi \to 2 \varphi$ $(2\varphi \to 2~{\rm SM})$
process is present in eq.\eqref{eq:phi}. The relative
contribution of these two processes in the evolution of $Y_\varphi$ is clearly
seen in this figure. If we consider only the sub-dominant 
$2 \varphi \to 2~{\rm SM}$ process, $\varphi$ freezes-out earlier (red dotted line)
due to small $\Gamma_{2\varphi \to 2~{\rm SM}}$, whereas, the
dominant $3\varphi \to 2\varphi$ process maintains $\varphi$ in thermal bath for 
longer duration (brown solid line).  
Thus the freeze-out abundance of 
$\varphi$ (blue dashed line) is governed mainly by the dominant 
$3 \varphi \to 2 \varphi$ process due to larger 
$\Gamma_{3\varphi \to 2\varphi}$ for our choice of model
parameters. Therefore, we can safely ignore the  
second term in eq.\eqref{eq:phi} and the modified Boltzmann equation takes the 
following form:
\begin{eqnarray}
 \nonumber
 \dfrac{dY_{\varphi}}{dx}&=&- 0.116 \frac{g_{s}^2}{\sqrt{g_{\rho}}} \dfrac{M_{\varphi}^4}{x^5} M_{pl} \left< \sigma v^2\right>_{3 \varphi \rightarrow 2 \varphi}(Y^3_{\varphi}-Y^2_{\varphi}Y^{eq}_{\varphi})\\
 &&-\sqrt{\frac{45}{4 \pi^3}} \left< \Gamma _{\varphi \rightarrow \chi \nu} \right> \frac{x}{M_{\varphi}^2} \frac{M_{pl}}{\sqrt{g_{\rho}}}Y_{\varphi} .
 \label{eq:phi_1}
 \end{eqnarray}
Based on the above argument we can identify the parameter space for 
scenario-I  satisfying the criteria: $x_F^{3\varphi \to 2\varphi}> x_F^{2\varphi \to 2{\rm SM}}$.  
In Fig.\ref{b2} we display the parameter space for this scenario in
$M_{\varphi}$ vs. $\lambda_{\varphi}$ plane with $\mu_{\varphi}/M_{\varphi}=0.1$ 
for two different values $\lambda_{\varphi H}=\{10^{-2},10^{-4}\}$ 
depicted by the blue and red shaded region respectively.
The criteria for scenario-I  holds only for the region left to individual lines. 
With an increase in $M_{\varphi}$ ,$\Gamma_{3\varphi \to 2 \varphi}$ becomes more mass suppressed compared to  $\Gamma_{2 \varphi \to 2 {\rm SM}}$.
Hence for fixed values of $\lambda_{\varphi},$ and $~\lambda_{\varphi H}$ with increasing $M_{\varphi}$, $\Gamma_{3\varphi \to 2 \varphi}$ falls below $\Gamma_{2 \varphi \to 2 {\rm SM}}$ and scenario-I doesn't hold anymore for the parameter space right to the colored lines. 
With an increase in $\lambda_{\varphi H}$, $\Gamma_{2 \varphi \to 2 {\rm SM}}$ increases and eventually $\Gamma_{3\varphi \to 2 \varphi}$ falls below $\Gamma_{2 \varphi \to 2 {\rm SM}}$ even with lower $M_{\varphi}$.
For that reason we see the shaded region moves toward lower $M_{\varphi}$ (towards left) with an increase in $\lambda_{\varphi H}$. 
The regions right to the colored lines demand a different treatment which will be discussed shortly.
 
Before we conclude this part of our analysis, 
it is worth noting that the present dark sector 
dynamics also allows $4 \varphi \to 2 \varphi $ 
number changing process involving the same $\lambda_\varphi$ coupling that
is responsible for $3\varphi \to 2\varphi$ process. Inspite of the same interaction
strength $(\lambda_\varphi)$, $4\varphi \to 2 \varphi$ process is more phase space 
suppressed compared to that of $3 \varphi \to 2 \varphi $, hence, we neglect it
in our numerical calculation of $Y_\varphi$.
\end{itemize}

\begin{itemize}
 \item {\bf Scenario II}: In this picture we consider 
$\Gamma_{2\varphi \to {\rm 2 SM}} \gg \Gamma_{3 \varphi \to 2 \varphi}$, which is
contrary to the previous scenario. In this case freeze-out of 
$\varphi$ is dictated by $2\varphi \to 2{\rm SM}$ annihilation process that 
keeps $\varphi$ in thermal bath for a longer period compared to 
${3 \varphi \to 2 \varphi}$ process.
Hence the freeze-out of $\varphi$ occurs at 
$x= x_{F}^{\rm tot} \approx x_F^{2\varphi \to 2{\rm SM}}>x_F^{3\varphi \to 2\varphi} $. 

\begin{figure}[tbh!]
\subfigure[\label{c1}]{
    \includegraphics[scale=0.39]{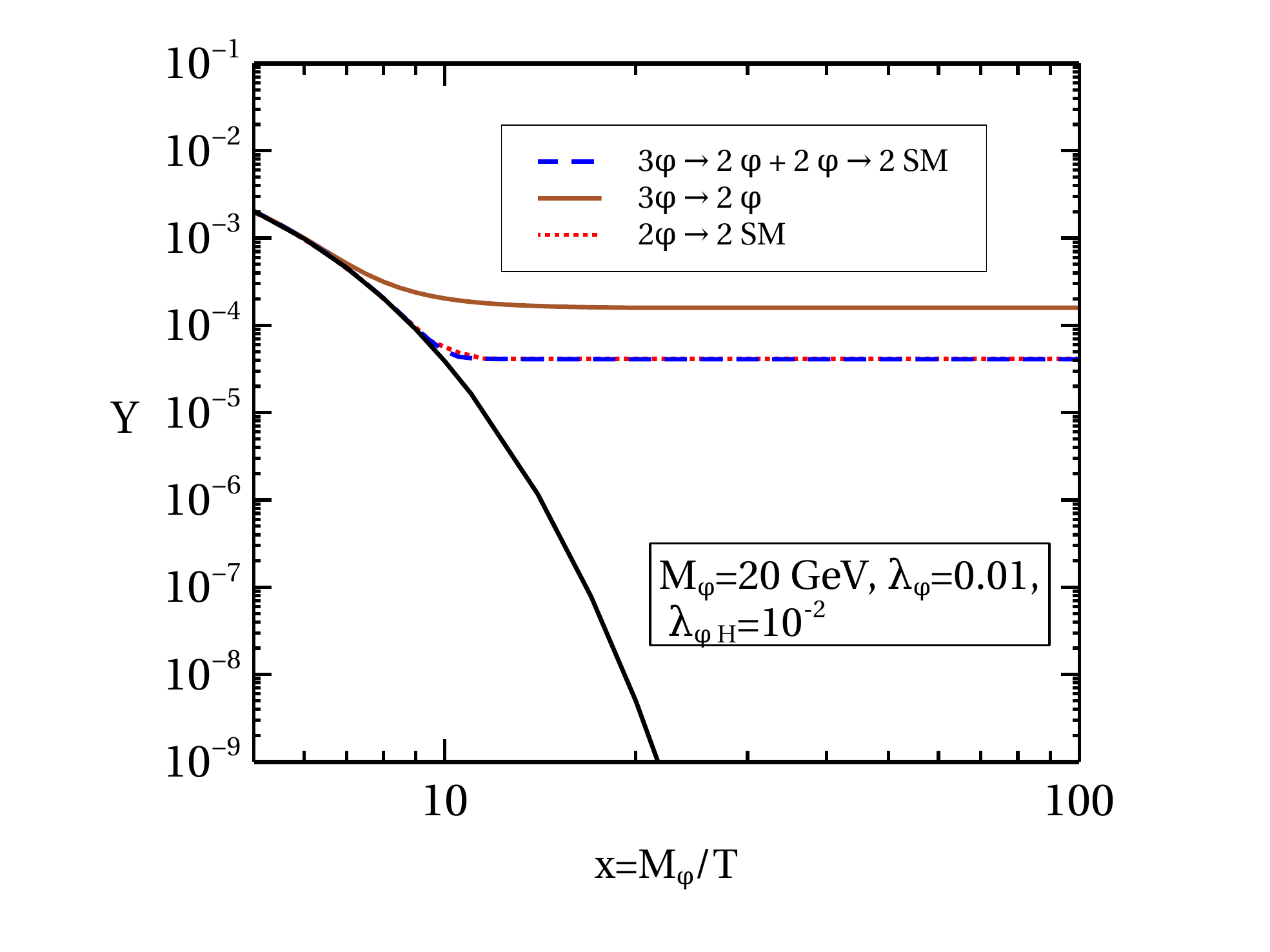}}
     \subfigure[\label{c2}]{
    \includegraphics[scale=0.39]{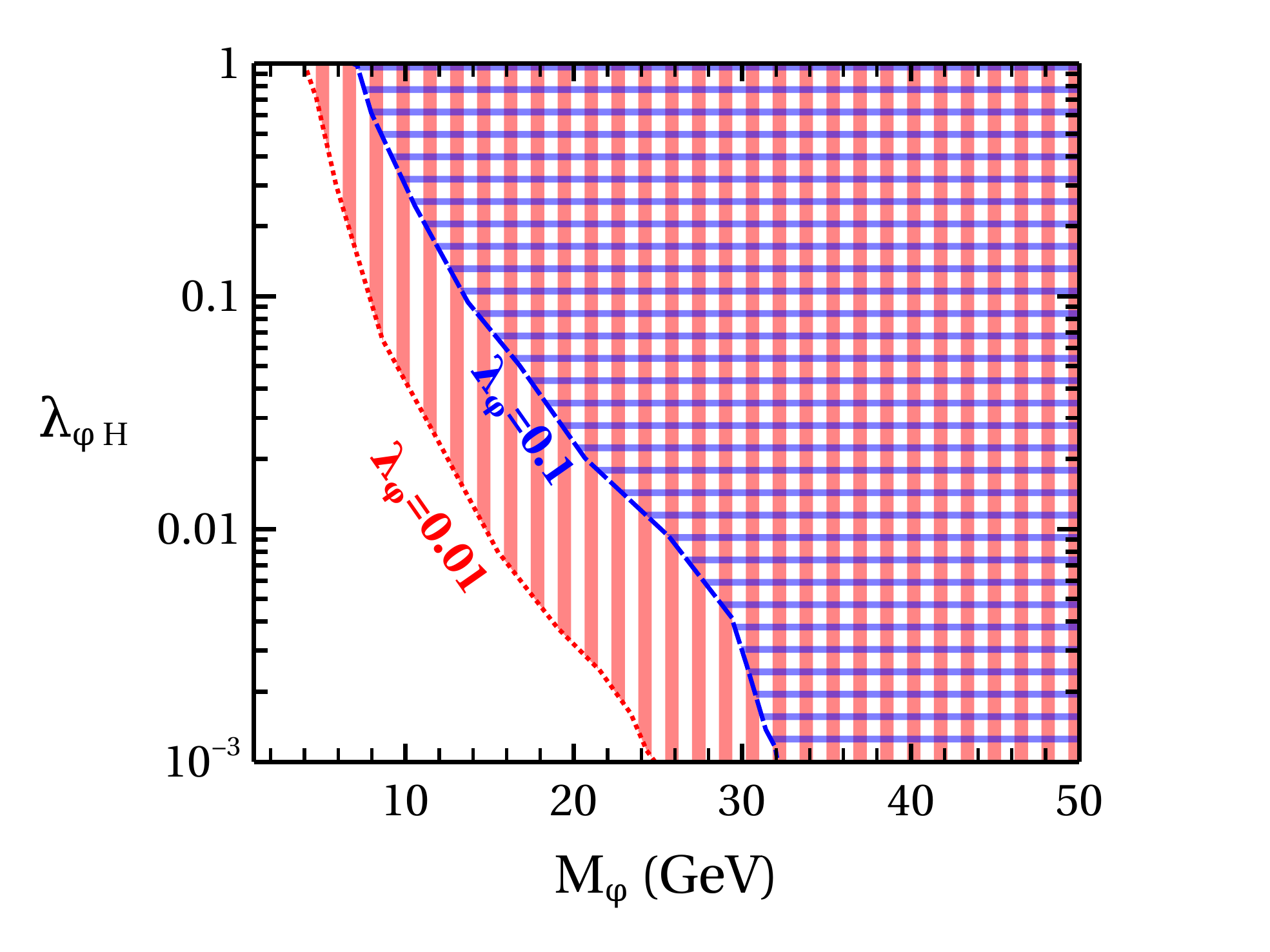}}
    \caption{{\it (a)Thermal freeze-out of $\varphi$ for $M_\varphi=20~ {\rm GeV},\mu_\varphi/M_\varphi=0.1,\lambda_{\varphi   H}=10^{-2},\lambda_\varphi=0.1$. The black solid line signifies $Y_{\varphi}^{eq}$.
    The blue dashed line corresponds to $Y_{\varphi}$ considering contributions from both $3 \varphi \to 2 \varphi$ and only $2\varphi \to 2~ {\rm  SM}$ processes. The brown solid line and red dotted line represent  $Y_{\varphi}$ considering only $3 \varphi \to 2 \varphi$ and $2\varphi \to 2~{\rm SM}$ process respectively.  (b)Parameter space for scenario-II in $M_{\varphi}$ vs.$\lambda_{\varphi}$ plane with $\mu_{\varphi}/M_{\varphi}=0.1$ for two different values of $\lambda_{\varphi H}=\{10^{-2},10^{-4}\}$ depicted by the blue and red shaded region respectively. The criteria for scenario-II  holds only for the region right to individual lines where $x_F^{\rm tot}$ is governed by  $2 \varphi \to 2~{\rm SM}$  process.}}
    \label{fig:3to2b}
\end{figure} 
 
In Fig.\ref{c1} we report the evolution 
of $Y_\varphi$ as a 
function of $x=\frac{M_\varphi}{T}$ for $\lambda_\varphi=0.01$ 
keeping other parameters same as 
in Scenario-I. From this figure it is evident that $Y_{\varphi}$ is entirely 
decided by $ 2 \varphi \to 2 {\rm SM}$ number changing processes contrary 
to the previous scenario where $3\varphi \to 2\varphi$ process was controlling
the dynamics. Thus eq.\eqref{eq:phi} can be simplified by neglecting 
the sub-dominant $ 3 \varphi \to 2 \varphi$ process:
\begin{eqnarray}
 \nonumber
 \dfrac{dY_{\varphi}}{dx}&=&- 0.264 \frac{g_{s}}{\sqrt{g_{\rho}}} \dfrac{M_{\varphi}}{x^2} M_{pl} \left< \sigma v^2\right>_{2\varphi\rightarrow 2{\rm SM}}(Y^2_{\varphi}-(Y^{eq}_{\varphi})^2)\\&&-\sqrt{\frac{45}{4 \pi^3}} \left< \Gamma _{\varphi \rightarrow \chi \nu} \right> \frac{x}{M_{\varphi}^2} \frac{M_{pl}}{\sqrt{g_{\rho}}}Y_{\varphi} ~ .
 \label{eq:phi_2}
 \end{eqnarray}

In Fig.\ref{c2} we display  parameter space for this scenario in
$M_{\varphi}$ vs. $\lambda_{\varphi H}$ plane with $\mu_{\varphi}/M_{\varphi}=0.1$ 
for two different values $\lambda_{\varphi}=(10^{-1}~\&~10^{-2})$ 
depicted by the blue and red shaded region respectively.
For the same reason discussed in the context of scenario-I,
in this case also $\Gamma_{3\varphi \to 2 \varphi}$ decreases with decrease in
 $\lambda_{\varphi }$
 and  finally falls below $\Gamma_{2 \varphi \to 2 {\rm SM}}$.
 And this phenomena is true even for lower $M_{\varphi}$. 
For this reason here also we see that the shaded region shifts 
towards lower $M_{\varphi}$ (left) with decrease in $\lambda_{\varphi}$. 
\end{itemize}

\noindent In summary the  main observation of this whole subsection are 
the following:
\begin{eqnarray}
{\rm Scenario ~I}&:& ~~\Gamma_{3\varphi \to 2\varphi} ~\gg ~\Gamma_{2\varphi \to 2{\rm SM}} ~~ \implies x_F^{\rm tot}~~ \approx~~ x_F^{3\varphi \to 2\varphi}>
x_F^{2\varphi \to 2{\rm SM}}~, \nonumber \\
{\rm Scenario~ II}&:& ~~\Gamma_{3\varphi \to 2\varphi} ~\ll ~\Gamma_{2\varphi \to 2{\rm SM}} ~~ \implies x_F^{\rm tot}~~\approx~~ x_F^{2\varphi \to 2{\rm SM}}>x_F^{3\varphi \to 2\varphi}.
\label{eq:criteria}
\end{eqnarray}

\noindent
It is worth mentioning that scenario-II is more common and has already been studied in different literature \cite{Ghosh:2022fws,Biswas:2022vkq}, where mother particles
are considered to have sizable annihilation cross-section with SM bath.
In this work our main focus is on scenario-I, although for the sake of completeness of the analysis we also discuss scenario-II. \\

{\bf{Step-II: Non thermal DM production}}

Following our previous discussion we now focus on the non-thermal 
production of DM ($\chi$) from the dilution of $\varphi$ density described 
by the last term in the R.H.S of eq.\eqref{eq:phi}. 
We solve the following Boltzmann equation to get the evolution 
of DM($\chi$) abundance,
\begin{equation}
  \dfrac{dY_{\chi}}{dx}=\sqrt{\frac{45}{4 \pi^3}} \left< \Gamma \right>_{\varphi \rightarrow \chi \nu} \frac{x}{M_{sc}^2} \frac{M_{pl}}{\sqrt{g_{\rho}}}Y_{\varphi} ,
  \label{eq:DM}
\end{equation}
where, $Y_{\chi}$ is the co-moving number density of DM $\chi$. 
In general, the solution of $Y_\varphi$ comes from eq.\eqref{eq:phi_1} for scenario-I  and in eq.\eqref{eq:phi_2} for scenario-II respectively.
In the calculation of $\Gamma(\varphi\to \chi \nu)$ we consider the Yukawa 
coupling $y_{1}$ in the range ($\sim 10^{-12}-10^{-15}$) so that the 
decay of $\varphi \to \chi + \nu $ happens in post BBN and pre CMB era. 
At this stage, we find
it worth discussing one subtle issue regarding the thermal averaged decay 
width $ \left< \Gamma \right>_{\varphi \rightarrow \chi \nu}$.
As we have pointed out before, that at the time when $\varphi$ freezes-out, 
it maintains the same temperature as the SM bath via the elastic 
scattering processes. {However the condition for kinetic equilibrium holds only upto freeze out and is no longer maintained at the time of decay($<T_{\rm BBN}$) as shown in Appendix \ref{sec:KEQ}.} 
This results the dark sector to acquire a different temperature $T'~(\neq T)$ 
than the thermal bath and this must be evaluated in order to 
get  $\left< \Gamma \right>_{\varphi \rightarrow \chi \nu}$.
In this work, as we are studying the dark sector 
dynamics at low temperature ($T' \ll M_{\varphi}$), and in this limit the 
thermally averaged decay width can simply be approximated as 
$\left< \Gamma \right>_{\varphi \rightarrow \chi \nu}(T') 
\approx \Gamma_{\varphi \rightarrow \chi \nu}$ \cite{Hall:2009bx,Biswas:2018ybc},
thus reducing the complication of tracking temperature dependence of 
the evolution of $\varphi$. After solving eq.\eqref{eq:DM} we get the 
complete picture of DM production as shown by red solid line in the left 
panel of Fig.\ref{fig:c2}.

\section{Light neutrino production before CMB}
\label{sec:neff}

Now we discuss the production of supplementary light neutrinos from the 
late time decay of $\varphi$ and the relevant mechanism of verifying those light degrees 
of freedom at CMB.
As revealed earlier, neutrinos that 
are produced after neutrino decoupling ($T\lesssim 2$ MeV) would
inject entropy in the neutrino bath.    
At the time of CMB, the number of relativistic neutrino degrees of freedom is expressed as, 

\begin{equation}
N_{\rm eff}^{\rm CMB} = \frac{8}{7} \left(\frac{11}{4} \right)^{4/3} \frac{\rho_{\nu}^{\rm SM}}{\rho_{\gamma}}\Bigg|_{\rm T=T_{CMB}},
\end{equation}
where, $\rho_{\nu}^{\rm SM}=3 \times 2\times \frac{7}{8} \times \frac{\pi^2}{30} (T_{\nu}^{\rm SM})^4$
and $\rho_{\gamma}=2 \times \frac{\pi^2}{30} T^4 $ are energy densities of neutrino and photon  respectively. Due to the extra neutrino injection from  the non-thermal decay of $\varphi$, the energy density of the neutrino bath 
increases to $\rho'_{\nu}$ ($\rho'_{\nu} >\rho^{\rm SM}_{\nu}$). In this case,
the relativistic neutrino degrees of freedom($N'_{\rm eff}$) also differs
from the prediction of SM at the time of CMB. 
We parameterise this deviation at the time of CMB in the following manner,
\begin{equation}
\rm \Delta N_{eff} = \left(\frac{\rho_{\nu}^\prime}{\rho_{\nu}^{\rm SM}}-1 \right)\,N_{\rm eff}^{\rm SM}\Bigg|_{\rm T=T_{CMB}}.
\label{eq:neff}
\end{equation}

\noindent We now solve the following Boltzmann equation to estimate the
evolution of ${\rm \rho_{\nu}^\prime}$ with temperature,
\begin{eqnarray}
\frac{d \rho'_\nu}{d x} = - \dfrac{4\,\beta \,\rho'_{\nu}}{x} + \frac{1}{x H (x)}\left< E \Gamma\right>_{\varphi \rightarrow \chi \nu} Y_{\varphi}~ s~,
\label{eq:rhonupr} 
\end{eqnarray}
where the term $\beta$
indicates the variation of $g_s(T)$ with T and is defined as
\begin{equation}
\beta (T)= 1+ \frac{1}{3}\frac{T}{g_{s}(T)} \frac{d\,g_{s}(T)}{dT}.
\end{equation}
where, $x$ is the dimensionless variable as mentioned earlier in context of equation \eqref{eq:phi}.  
$ Y_{\varphi}$ is the co-moving number density of $\varphi$ which is computed  by solving eq.\eqref{eq:phi_1} or eq.\eqref{eq:phi_2} depending on the scenario we consider. 
$g_s(x)$ is the number of effective degrees of freedom related to the entropy density and $s$ is the co-moving entropy density.
The term {$\left<E \Gamma \right>_{\varphi \rightarrow \chi \nu}$} in eq.\eqref{eq:rhonupr} is the most crucial ingredient  in this analysis which represents the 
thermal averaged energy density transferred to neutrino sector and is defined 
as 
\begin{eqnarray}\nonumber
{\small \left< E \Gamma\right>_{\varphi \rightarrow \chi \nu}=  \frac{|\mathcal{M}|_{\varphi \rightarrow \chi \nu}^2}{32\pi} \frac{{\left(M_{\varphi}^2-M_{\rm DM}^2\right)}}{M_{\varphi}^2} \left(1-\frac{M_{\rm DM}^2}{M_{\varphi}^2}\right)}.
\label{eq:rhonu}
\end{eqnarray}

The first term in the R.H.S of  eq.\eqref{eq:rhonupr} is responsible for the 
dilution of $\rho_{\nu}'$ due to expansion of the universe
while the second term decides the evolution of augmented contribution to 
$\rho'_{\nu}$ from $\varphi$ decay.
The evolution of $\rm{\rho_{\nu}^{SM}}$ after the decoupling of neutrinos 
is governed by only the expansion effect.
Thus in the absence of any new source $\rm{\rho_{\nu}^{SM}}$ can be computed by setting the second term of the R.H.S of eq.\eqref{eq:rhonupr} equal to zero and considering only the dilution of energy density.
\section{Relic density and $\mathbf{\Delta N_{\rm eff}}$}
\label{sec:relic_neff}
So far we have built up the basic framework of the underlying dynamics of dark sector 
particles ($\varphi$ and $\chi$) that provided freeze-in DM as well as yielded 
extra active light neutrino that with its possible footprints in
$\Delta \rm N_{ eff}$.
In this section, we perform an exhaustive numerical analysis of 
Scenario-I and Scenario-II to quantitatively estimate  
phenomenological consequences of the late-time decay of $\varphi$
in the light of current and future measurements of $\Delta \rm N_{eff} $. 
For this we first scrutiny the dependence of 
DM relic density and the $\Delta \rm N_{ eff}$
on various model parameters
 as elaborated 
in sec-\ref{sec:dynamics} and sec-\ref{sec:neff} respectively.   
\subsection{Relic density}
To calculate the DM relic density, we numerically solve  eq.\eqref{eq:DM} along with either eq.\eqref{eq:phi_1} (for scenario-I) or eq.\eqref{eq:phi_2} (for scenario-II). The solution of the coupled Boltzmann equations for each scenario yields $Y_\varphi$ and $Y_\chi$ as a function of $x$($=M_\varphi/T$). Using the co-moving density of DM, $Y_\chi$ at $x\to \infty$, one finds out DM relic density as \cite{Kolb:1990vq}:
\begin{equation}
  \Omega_\chi h^2= 2.755\times 10^8 \times \left(\frac{M_{\rm DM}}{\text{GeV}} \right) \times Y_{\chi}^{\rm today} , 
  \label{eq:omega}
\end{equation}
where $Y_{\chi}^{\rm today}=Y_\chi(x\to \infty)$. 
The precise determination of $Y_{\chi}^{\rm today}$ is highly model dependent.
In the following two sub-sections we pin down 
$Y_{\chi}^{\rm today}$ for scenario-I and II and  
corresponding relic densities.

\begin{figure}[tbh]
\centering
\subfigure[\label{l1}]{
\includegraphics[scale=0.32]{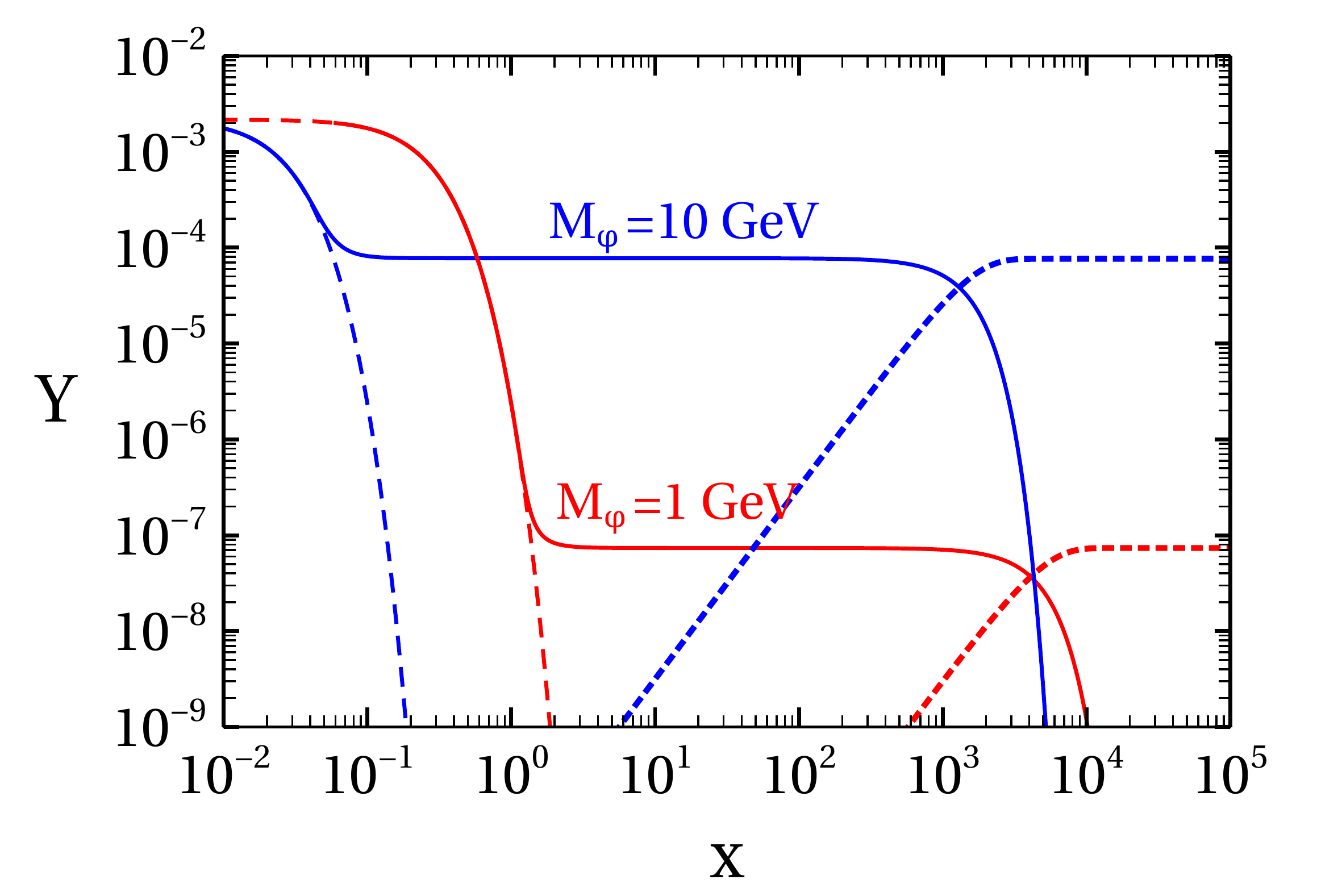}} 
\subfigure[\label{l2}]{
\includegraphics[scale=0.32]{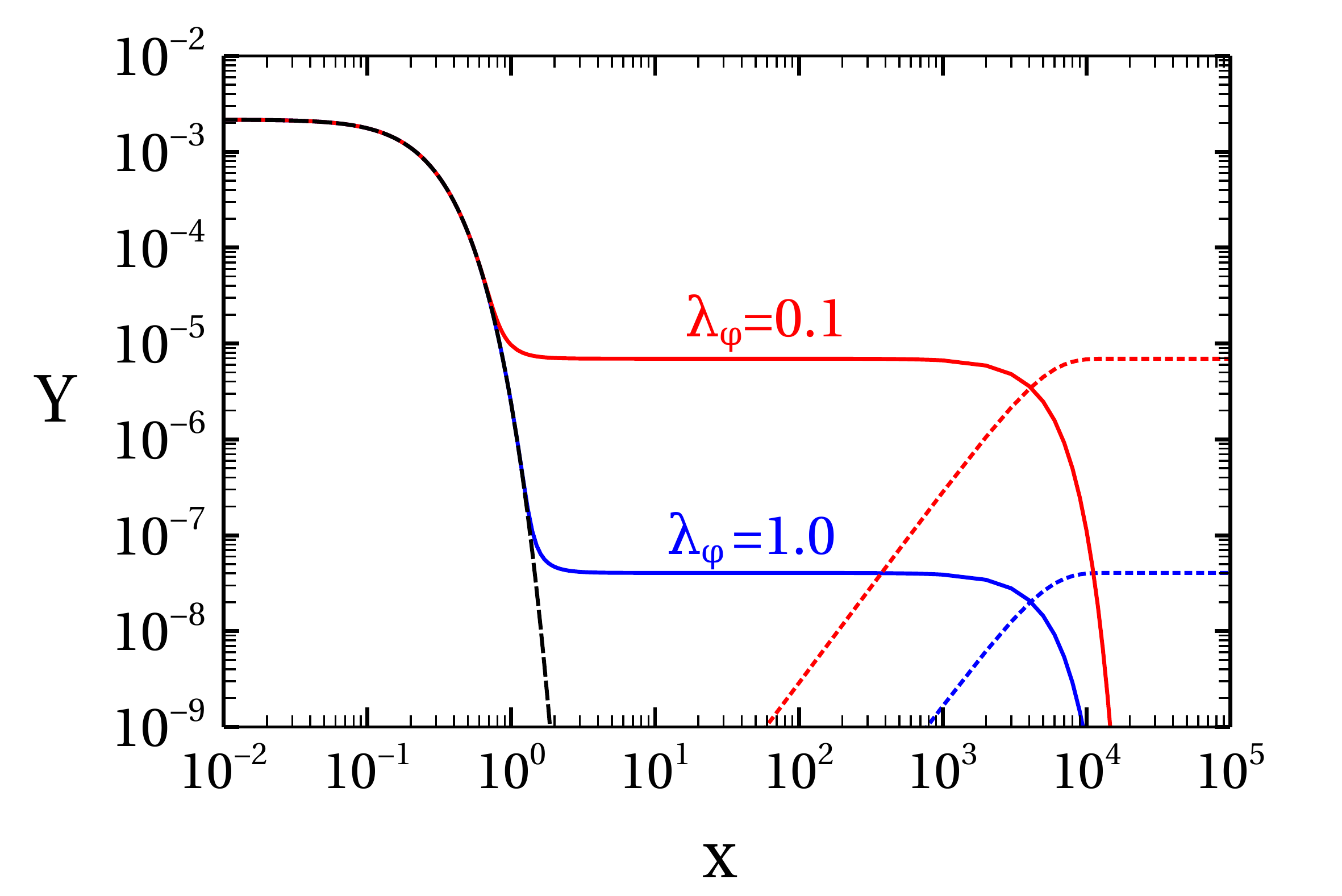}} 
 \caption{\it Evolution of co-moving abundances of $\varphi$ (solid line) and DM($\chi$) (dotted line) with $x(\equiv1/T)$ ($T$ in GeV) for scenario-I. In (a)for a fixed  $\lambda_{\varphi}=1.0$ with two different values of $M_{\varphi}$ and in (b) for a fixed $M_{\varphi}=1$ GeV with two different values of $\lambda_{\varphi}$ are shown. The Higgs portal coupling is considered here to small, $\lambda_{\varphi H}=10^{-4}$ in order to realise the scenario. The other parameters like  $y_1=10^{-12}$, $M_{\rm DM}=400$ keV and $\mu_\varphi/M_\varphi=0.1$ are kept same for both plots.}
\label{fig:ydm}
\end{figure}

\underline{\bf Scenario-I}:
As shown before, densities of $\varphi$  and $\chi$ for the scenario-I 
are mainly driven by $3\varphi \to 2\varphi$ number changing 
process in the dark scalar sector.
Based on this number changing process, we calculate the co-moving abundances
of $\varphi$ and $\chi$ and show their evolution with $x(=M_\varphi/T)$ 
in Fig.\ref{fig:ydm}. The solid, dashed and dotted lines 
signify $Y_{\varphi},\, Y_{\varphi}^{eq}$ and $Y_{\chi}$ respectively. 
It can be seen from these figures that the late-time decay of 
$\varphi$ (solid lines) produces the abundance of $\chi$. 
As the $\varphi \to \chi + \nu $ decay proceeds, the number density 
of $\varphi$ slowly changes into $\chi$ number density and eventually
at the end of the decay, the density of $\varphi$ completely dilutes to 
$\chi$ number density ($Y_{\chi}\equiv Y_{\varphi}$ at $\tau\gg \tau_{\varphi}$).
One can easily understand this from the fact that $\varphi \to \chi \nu$ decay
is the only possible decay mode of the NLP $(\varphi)$
\cite{Coy:2021sse}. Thus, in the generation of $Y_\chi$ from $Y_\varphi$, 
the magnitude of the Yukawa coupling $(y_{_1}>0)$ has hardly any role to play, except for 
setting the lifetime of $\varphi$ and this provides $Y_\varphi\big({x_F^{3\varphi 
\to 2\varphi}} \big) \simeq Y_\chi^{\rm today}$. Therefore the relic density of DM given in 
eq.\eqref{eq:omega} turns out to be $\Omega_\chi h^2 \propto M_{\rm DM} 
\times Y_{\varphi}\big({x_F^{3\varphi \to 2\varphi}} \big)$.
Consequently in order to get 
fixed $\Omega_{\chi}h^2$, any increase in $M_{\rm DM}$
demands a decrease in $Y_{\varphi}\big({x_F^{3\varphi \to 2\varphi}} \big)$ and vice versa.   
We have pointed out before that 
$\varphi \to \chi ~\nu$ decay to happen between BBN and CMB the value of $y_1$ should lie in the range : $\{10^{-12} - 10^{-15}\}$.
As a sample representative value we set $y_{_1}=10^{-12}$ throughout our numerical analysis.
To show the evolution of $Y_\varphi$ and $Y_\chi$ with temperature, we fix $\mu_\varphi/M_\varphi=0.1$ and $M_{\rm DM}=400$ keV for both plots. 
We consider $\lambda_{\varphi H}= 10^{-4}$ to realize  Scenario-I.

In Fig.\ref{l1} we present evolution of densities $Y_\varphi$ (solid line)
and $Y_\chi$ (dotted line) as a function of dimension less parameter 
$x(= M_\varphi/T)$ for two different $M_{\varphi}$ and a fixed 
self-interaction coupling $\lambda_{\varphi}=1.0$. The dynamics of dark sector particles for $M_{\varphi}=1$ GeV and $M_{\varphi}=10$ GeV are depicted by red and blue colors respectively.
With the increase of $M_{\varphi}$, $\left<\sigma v^2\right>_{3\varphi \to 2\varphi}$ encounters phase space and propagator  suppression which is also understood from the expression given in appendix \ref{sec:3DMto2DM}.
As $Y_\varphi$ goes like $Y_\varphi \propto 1/\left< \sigma v^2 \right>_{3\varphi \to 2\varphi}$ (using analytical solution \cite{Bhattacharya:2019mmy}), with the smaller  $\left<\sigma v^2\right>_{3\varphi \to 2\varphi}$, the thermal freeze-out of $\varphi$ happens at earlier time with higher abundance $Y_{\varphi}$ and eventually this $Y_\varphi$ is transfered to the $Y_\chi$. As a result $Y_{\chi}$ is also higher for higher $M_{\varphi}$. This feature is portrayed in the Fig.\ref{l1} where higher(lower) value of $M_{\varphi}$ leads to the higher(lower) abundance $Y_{\chi}$ represented by the red(blue) dotted line.

To study the role of dark scalar self-coupling,$\lambda_\varphi$ on DM abundance,
 in Fig.\ref{l2} we show the variation in $Y_{\chi}$ 
for two different values of $\lambda_{\varphi}=$  $0.1$ (red line) and $1.0$ 
(blue line) keeping $M_{\varphi}=1$ GeV. It is obvious, that 
as the value of $\lambda_{\varphi}$ increases, the thermal averaged 
cross-section $\left<\sigma v^2\right>_{3\varphi \to 2\varphi}$ also increases 
which eventually reduces the abundance $Y_{\varphi}$
and finally this reduced $Y_{\varphi}$  generates 
lower $Y_{\chi}$. This is elucidated in Fig.\ref{l2}, where a 
higher(lower) value of $\lambda_{\varphi}$ gives a lower(higher) 
$Y_{\chi}$ as it is shown by the blue(red) dotted line.    

The other parameter $\mu_{\varphi}$ with mass dimension,  is also responsible for $3\varphi \to 2\varphi$ processes as $\left<\sigma v^2 \right>_{3\varphi \to 2\varphi}\propto (\mu_\varphi/M_\varphi)^2$ (see Appendix \ref{sec:3DMto2DM}).
With an increase in the ratio $\mu_{\varphi}/M_\varphi$, the cross-section will enhance 
leading to a decrease in $Y_\varphi$ as well as $Y_{\chi}$.
For simplicity,  we consider the ratio $\mu_{\varphi}/M_\varphi=0.1$ throughout our analysis and is consistent with the theoretical upper bound on $\mu_{\varphi}$ coming from stable vacuum as discussed in Appendix \ref{sec:stab}. 

\begin{figure}[tbh!]
\centering
\includegraphics[scale=0.6]{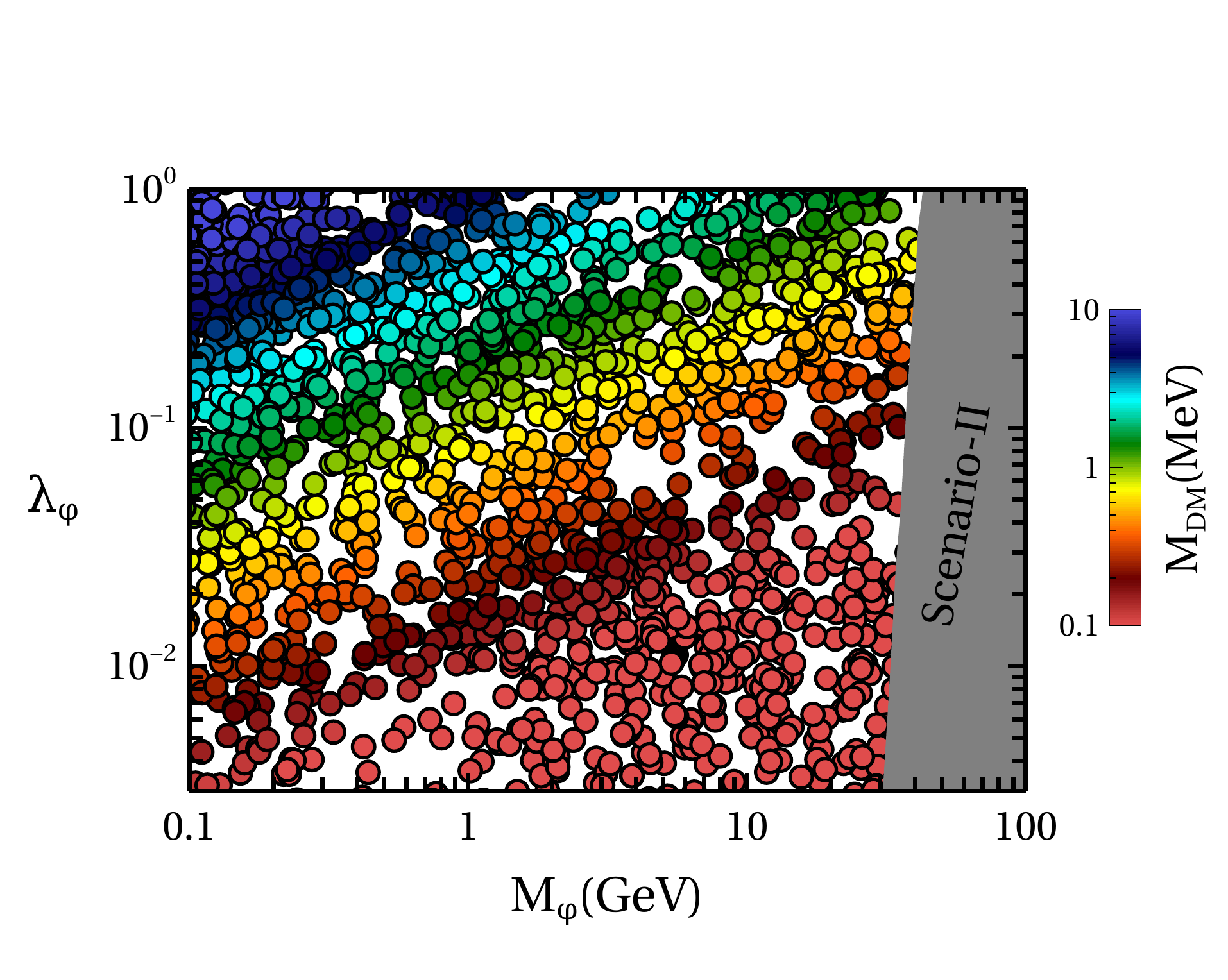} 
 \caption{\it DM relic density satisfied points in $M_\varphi-\lambda_\varphi$ plane  for scenario-I with $\mu_{\varphi}/M_{\varphi}=0.1 $, $\lambda_{\varphi H}=10^{-4}$ and $y_1=10^{-12}$. The color gradient indicates the range of $M_{\rm DM}$ satisfying the correct relic density. The shaded region corresponds to the parameter space where scenario-II is dominating over scenario-I. White regions are just computational artifact associated with the scan.}
\label{fig:relic}
\end{figure}

After describing the dependence of relic abundance on different model
parameters, we now present the allowed region of dark sector parameter
space from DM observed density, 
($\Omega_{\rm DM}h^2=0.120\pm0.001$) given by PLANCK \cite{Planck:2018vyg}. 
We perform
a numerical scan on the model parameters in the following range
\begin{eqnarray}
~M_\varphi~:~\{0.1~-100~{\rm GeV} \}, ~~~\lambda_\varphi~:~~\{0.001-1\}~;
\end{eqnarray}
to calculate $\Omega_{\chi} h^2$ using eq.\eqref{eq:omega}.
We keep  other parameters fixed as in Fig.\ref{fig:ydm}.
and to allow the on shell decay $\varphi\to \chi +\nu$ we set $M_\varphi > M_{\rm DM}$.

Our scan result is displayed in Fig.\ref{fig:relic}, where we show  
points satisfying relic density constraints in  $\lambda_{\varphi}$ vs. $M_{\varphi}$ plane. 
The grey shaded region corresponds to the parameter space where scenario-II dominates which demands a different analysis and will be discussed shortly.
The color gradient in the above figure represents DM mass range varying from 0.1 MeV to 10 MeV set by the observed relic density constraint. 
One can see from this figure that the higher value of $\lambda_{\varphi}$ prefers to higher value of $M_{\rm DM}$.
As explained in the context of Fig.\ref{fig:ydm}, with the increase of $\lambda_{\varphi}$, $Y_\chi$ decreases and hence higher value of
$M_{\rm DM}$ is required in order to satisfy the correct relic density as depicted in above Fig.\ref{fig:relic}. 
For simplicity we restrict our scan within the specified range of $\lambda_{\varphi}$ mentioned above.
However one can make the scan even for higher value of $\lambda_{\varphi}\gtrsim 1.0$ within the perturbativity limit. For those values of $\lambda_\varphi$ even heavier DM mass,($10~{\rm MeV}\lesssim M_{\rm DM} < M_\varphi$) will be allowed by the relic density constraint.


\underline{\bf Scenario-II}:
We shall now move to the second scenario where the density of $\varphi$ is mainly driven by $2\varphi \to 2  ~{\rm SM}$ number changing process.
Following our earlier discussion we know that 
the density of $\varphi$ converts into the density of DM \big($Y_\varphi\big({x_F^{2\varphi \to 2{\rm SM}}} \big) \simeq Y_\chi^{\rm today}$\big) via the late time decay of $\varphi$. 
Therefore the relic density of DM given in eq.\eqref{eq:omega} becomes  $\Omega_\chi h^2 \propto M_{\rm DM} \times Y_{\varphi}\big({x_F^{2\varphi \to 2{\rm SM}}} \big)$. 
Similar to previous scenario $Y_\varphi$($x_F^{2\varphi \to 2{\rm SM}}$), decreases with increase in $M_{\rm DM}$ in order to get fixed density and vice versa.
One can also analytically express the yield of $\varphi$ at freeze-out as: $Y_\varphi \propto 1/\langle \sigma v \rangle_{2\varphi \to 2{\rm SM}}$ \cite{Kolb:1990vq}. 
For heavier $M_\varphi$, more annihilation processes of $\varphi$ to SM pairs kinematically open up and enhance the cross-section. 
Thus $\langle \sigma v \rangle_{2\varphi \to 2{\rm SM}}$ can be expressed as  $\sum_{X=SM}\langle \sigma v \rangle_{\varphi \varphi \to {X \overline{X}}}~ \Theta\big(M_\varphi-M_X\big)$ where $\Theta$ is the Heaviside step function.   
For a fixed $M_\varphi$,  $Y_\varphi$ as well as $Y_\chi$ decreases as one increases $\lambda_{\varphi H}$ since $\langle \sigma v \rangle_{2\varphi \to 2{\rm SM}} \propto \lambda_{\varphi H}^2$. 
However, here the dependence of $Y_{\chi}$ on $M_{\varphi}$ is contrary to that of scenario-I.
In this case with the increase in $M_{\varphi}$, the annihilation cross-section, $\langle \sigma v \rangle_{2\varphi \to {\rm SM}}$ also increases 
for the aforementioned reasons and thus
resulting a decrease in $Y_{\varphi}$ as well as $Y_{\chi}$ \cite{Ghosh:2022fws}.

Now in order to find a consistent parameter space satisfying  observed relic density measured by PLANCK\cite{Planck:2018vyg}, we perform
a numerical scan of the relevant parameters for scenario-II 
in the following range:
\begin{eqnarray}
\label{eq:SII_scan}
~M_\varphi~:~\{10-100~{\rm GeV} \},~~
~\lambda_{\varphi H}~:~\{10^{-3}-10^{-1} \}~;
\end{eqnarray}
whereas the other parameters are kept fixed as $\mu_\varphi/M_\varphi=0.1,~\lambda_{\varphi}=0.1~{\rm and}~y_1=10^{-12}$.  
The choices of dark sector parameters in eq.\eqref{eq:SII_scan} ensure that $ \Gamma_{2\varphi \to 2{\rm SM}} \gg  \Gamma_{3\varphi \to 2\varphi} $ which is required for the scenario-II.
We consider $M_\varphi$ up to $100$ GeV, beyond that $Y_\varphi$ is more suppressed resulting in a negligible contribution to $\rm \Delta N_{ eff}$ which will be discussed in due course of time.
\begin{figure}
    \centering
    \includegraphics[scale=0.6]{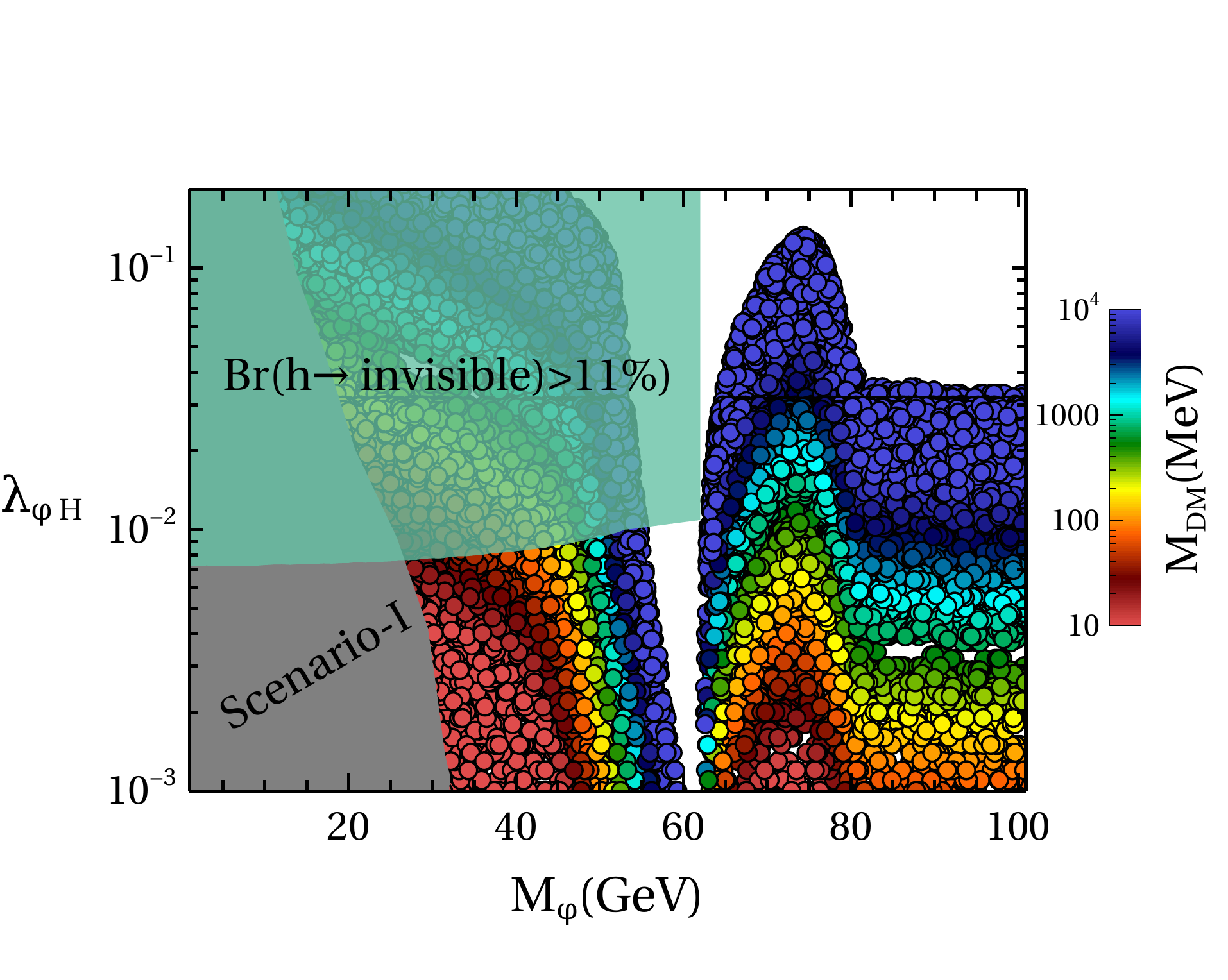}
    \caption{\it DM relic density satisfied points for scenario-II are shown in the $\lambda_{\varphi H}$ vs. $M_{\varphi}$ plane with $\lambda_{\varphi}=0.1$, $\mu_{\varphi}=0.1 M_{\varphi}$, $y_1=10^{-12}$ and the color gradient represents the variation in $M_{\rm DM}$ satisfying the correct relic density. The gray shaded region corresponds to the parameter space where scenario-I is dominating.}
    \label{fig:relic_sm}
\end{figure}

In Fig.\ref{fig:relic_sm} we plot correct relic density satisfied points in the $M_{\varphi}$ vs. $\lambda_{\varphi H}$ plane.
The variation of color gradient represents the variation of $M_{\rm DM}$ considered here. The correct relic density constraint sets the DM mass in the range $M_{\rm DM}$: $\sim \{10~{\rm MeV}-10~{\rm GeV} \}$ for our chosen parameters.
The gray shaded region on lower left corner of the above figure represents the region where scenario-II does not work paving way to Scenario-I.
With increase in $\lambda_{\varphi}(>0.1)$, Scenario-I will start to dominate over scenario-II even with higher value of $M_{\varphi}$ and the shaded region will move towards right accordingly.
For $M_\varphi < m_h/2$ , $h\to \varphi \varphi^*$ decay opens up
and contributes to the SM Higgs invisible decay width ($\Gamma_h^{\rm inv.}$) which is very precisely measured by CMS \cite{CMS:2018yfx}. The bound from $\Gamma_h^{\rm inv.}$ (discussed in Appendix \ref{sec:higgs}) excludes a significant part of the parameter space as shown by the light cyan region in Fig.\ref{fig:relic_sm}.    
As understood from the figure, scenario-II works in the higher range of $M_{\varphi}$ and the moderate values  $\lambda_{\varphi H}$ leading to lower $Y_{\chi}$ as discussed earlier in this subsection.
The $2\varphi\to2{\rm SM}$ annihilation cross-section near Higgs pole, $M_\varphi \sim m_h/2$, causes further suppression in $Y_{\chi}$. For $M_{\varphi}>M_{W}$, more final states 
open up resulting in even larger $\langle \sigma v \rangle_{2\varphi \to 2{\rm SM}}$. 
Thus to satisfy the observed DM density one has to reduce $\lambda_{\varphi H}$ in that 
region as shown in top right corner (white area) of the figure.
Therefore the scenario-II allows higher DM mass to satisfy the correct relic  density upto few GeV. 

\subsection{Contribution to $\mathbf{\Delta \rm N_{\rm eff}}$ at CMB }
In earlier sections, we have established that our main thrust of this whole
exercise is to calculate contributions to 
$\Delta {\rm N_{eff}}$ by extra active neutrinos produced in 
association with FIMP like DM from the late time decay of
a self interacting dark scalar $\varphi$. 
Simultaneously, we have also emphasized the possibility of 
correlating the dark matter mass with
the measured value of $\Delta {\rm N_{eff}}$. Thus, any 
precise determination of $\Delta {\rm N_{eff}}$ would provide  
an indirect probe of the dynamics of dark sector ivolving
a strongly self interactiong particle $\varphi$ as well as FIMP 
like DM. 
Based on our discourse in sec-\ref{sec:neff} we will now investigate 
dependence of dark sector model parameters in $\Delta \rm N_{eff}$ which 
is completely determined by the ratio $\rho'_{\nu}/\rho_{\nu}^{\rm SM}$.

\begin{figure}[h]
\centering
\subfigure[\label{r1}]{
\includegraphics[scale=0.3]{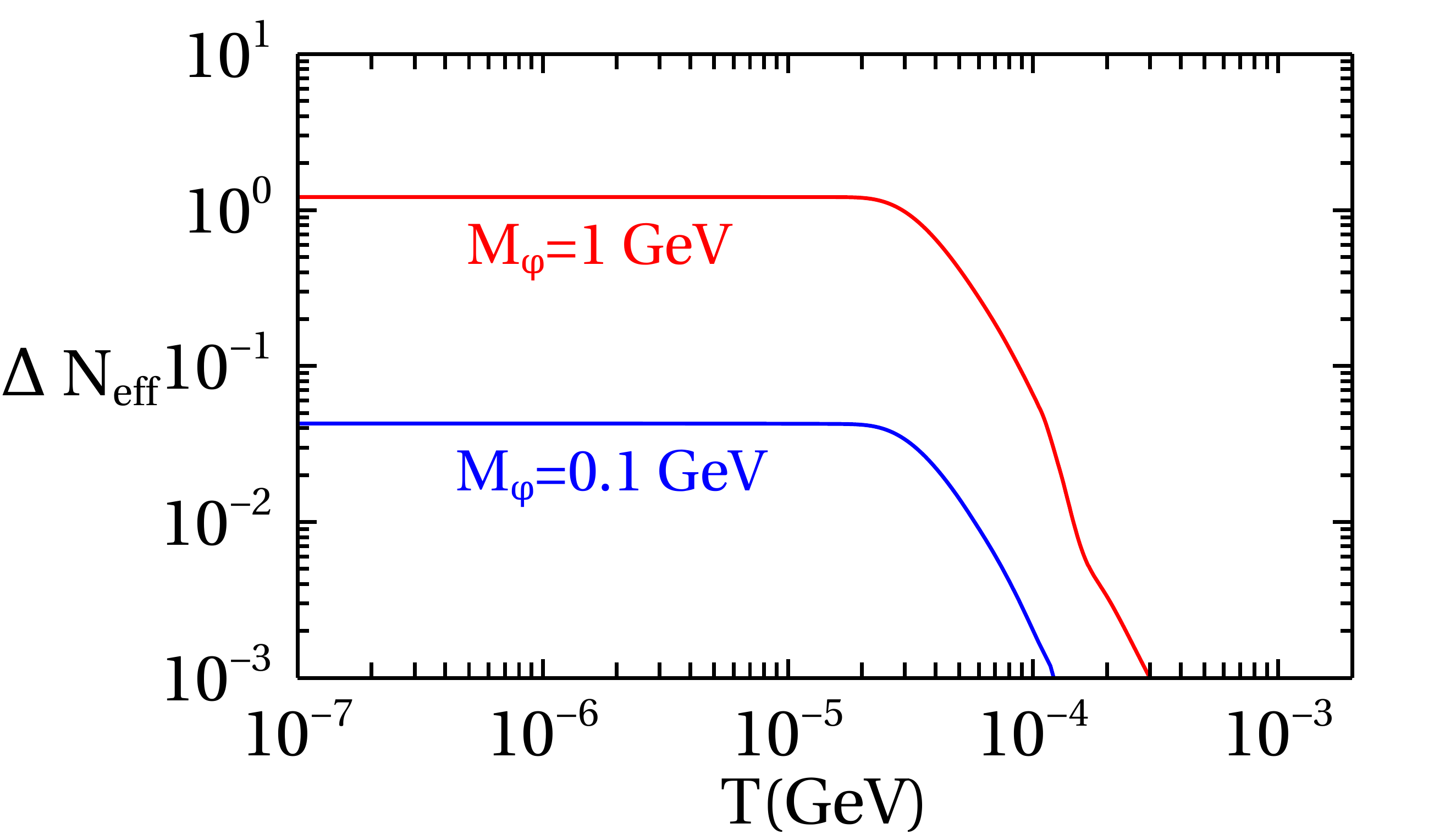}}
\subfigure[\label{r2}]{
\includegraphics[scale=0.3]{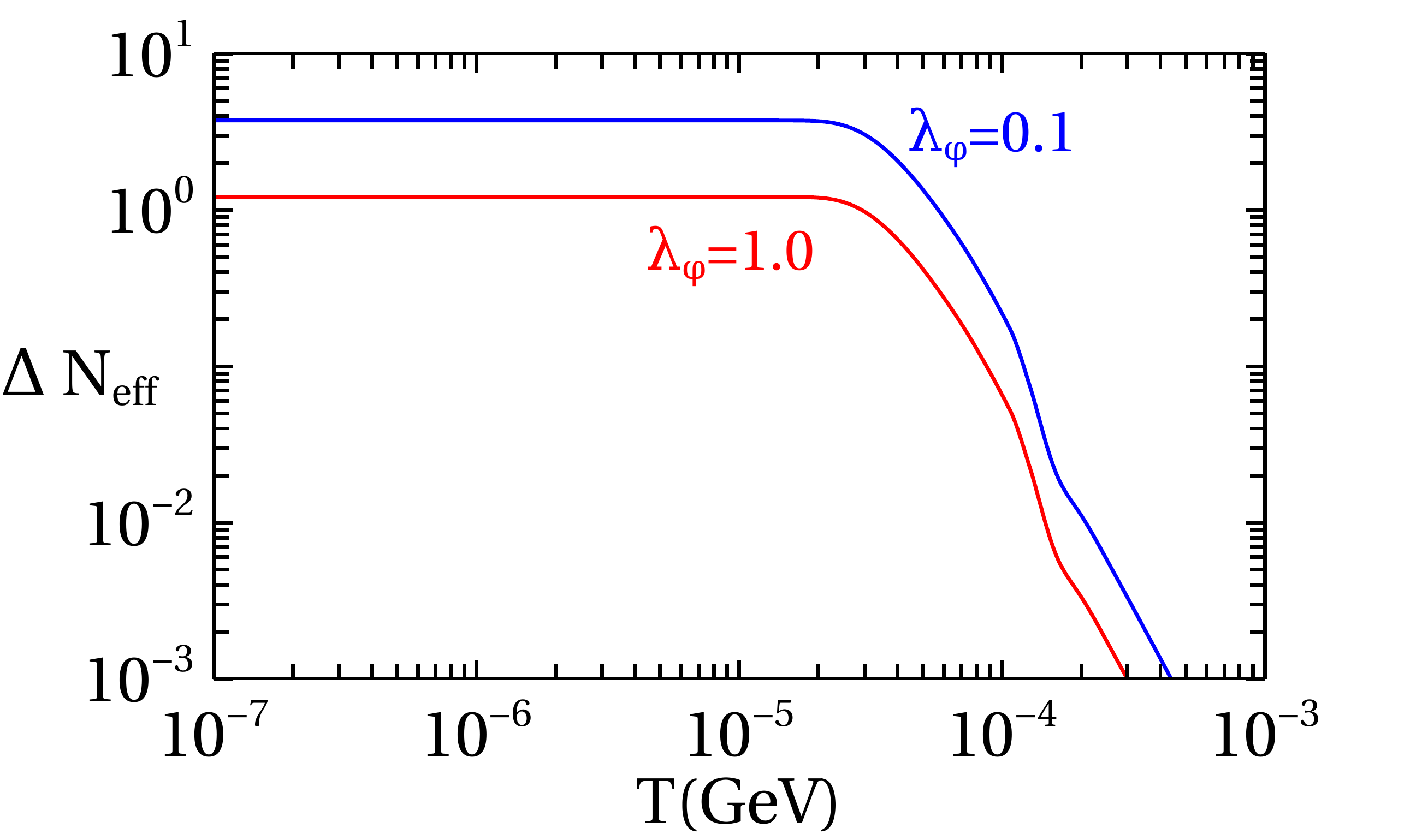}}
 \caption{\it Evolution of $\Delta N_{\rm eff}$  with temperature(T) for two different values of $ M_{\varphi}$ keeping $\lambda_{\varphi}=1.0$ fixed in (a) and for two different values of $\lambda_{\varphi}$ keeping $M_{\varphi}=1$ GeV fixed in (b) for scenario-I. Other parameters are kept fix as $y_1=10^{-12}$ and $M_{\rm DM}=400$ keV.}
\label{fig:ratio}
\end{figure}

In Fig.\ref{fig:ratio} we show the evolution of $\Delta N_{\rm eff}$ with 
temperature T for different set of model parameters as shown in the figure caption. 
We first numerically evaluate $\rho'_\nu /\rho_\nu^{\rm SM}$ by solving 
eq.\eqref{eq:rhonu} along with eq.\eqref{eq:phi_1} and then 
plug it into eq.\eqref{eq:neff} to estimate $\Delta \rm N_{eff}$.
From both the figures Fig.\ref{r1} and \ref{r2} we notice that at high $T$,
the $\Delta N_{\rm eff}$ is almost negligible 
because the entropy injection to neutrino bath is very small during the earlier epoch of $\varphi \to \chi \nu$ decay.
With the decrease in temperature, $\varphi$ freezes out from the thermal bath 
and decays into $\chi +\nu$ after BBN, generating a new source of 
active neutrinos that inject extra energy density to neutrino bath. 
This added neutrino density causes continuous growth 
of $\Delta {\rm N_{eff}}$ with lowering of temperature. With 
further decrease in the temperature, at some 
point $\varphi$ decay is completed and any auxiliary neutrino production also stops. 
Thus no more supplementary energy transfer to neutrino bath takes place 
and the ratio $\rho'_{\nu} / \rho_{\nu}^{\rm SM}$
attains its maximum possible value at that temperature. 
After that both $\rho'_{\nu}$ and 
$\rho_{\nu}^{\rm SM}$ dilutes in the same fashion with further decrease in 
temperature, resulting in a fixed ratio $\rho'_{\nu} / \rho_{\nu}^{\rm SM}$ 
which corresponds to a constant value of $\Delta N_{\rm eff}$. 
Since $\rho_{\nu}' \propto Y_{\varphi}$ (following eq.\eqref{eq:rhonupr}), higher value of $Y_{\varphi}$ leads
to higher energy transfer to neutrino bath  resulting in larger $\Delta \rm N_{eff}$ and vice-versa. 
We plot the evolution 
of $\Delta N_{\rm eff}$  in Fig.\ref{r1} 
for $M_{\varphi}=1$ GeV (red line) and  $M_{\varphi}=0.1$ GeV (blue line)
keeping $\lambda_{\varphi}=1.0$ fixed. 
In Fig.\ref{r2} we show the similar plot as in 
Fig.\ref{r1} but this time for a fixed $M_\varphi = 1 $ GeV and 
taking two values of $\lambda_{\varphi}=1.0$ (red line) and 0.1 (blue line). 
While generating these two plots, we fix $M_{\rm DM} = 400 $ MeV, 
and $y_1 = 10^{-12}$. 
The behavior of $\Delta \rm N_{eff}$ 
with the model parameters ($M_\varphi,~\lambda_\varphi$ and $\mu_\varphi/M_\varphi$) is same as of $Y_\varphi$ 
as discussed earlier for scenario-I (Fig.\ref{l1} and \ref{l2}) and the same dependence is depicted in  
 Fig.\ref{r1} and \ref{r2}.

The effect of $\lambda_{\varphi H}$ on $\Delta \rm N_{\rm eff}$ in 
Scenario-II is similar to Scenario-I. Following our previous argument, 
for any increase in the value of Higgs portal coupling $\lambda_{\varphi H}$,
$\varphi$ number density $Y_\varphi$ decreases and that leads to a diminished
contribution of active neutrinos in $\Delta \rm N_{eff}$. 
However, $M_\varphi$ dependence of 
$\Delta \rm N_{eff}$ shows opposite behaviour in Scenario-II than Scenario-I. 
Here, $\Delta \rm N_{eff}$ decreases with an increase in $M_\varphi$. 
The reason for this contrary nature follows the same argument as we 
revealed in the context of relic density calculation. For 
heavier $M_{\varphi}$, due to enhanced phase space one gets larger 
$\langle \sigma v \rangle_{2\varphi \to 2{\rm SM}}$ that leads to lower $Y_{\varphi}$ 
and finally lower $\Delta \rm N_{\rm eff}$. 
Hence the energy transferred to neutrino sector is too less to contribute significantly in $\Delta \rm N_{\rm eff}$ and
 the $\Delta \rm N_{\rm eff}$ for scenario-II will be far below the sensitivity of the current and future generation experiments.
In this paper we do not display the explicit parameter dependence in $\Delta \rm N_{\rm eff}$ in scenario-II, however similar study could be found in \cite{Ghosh:2022fws}.

\begin{figure}
    \centering
    \includegraphics[scale=0.6]{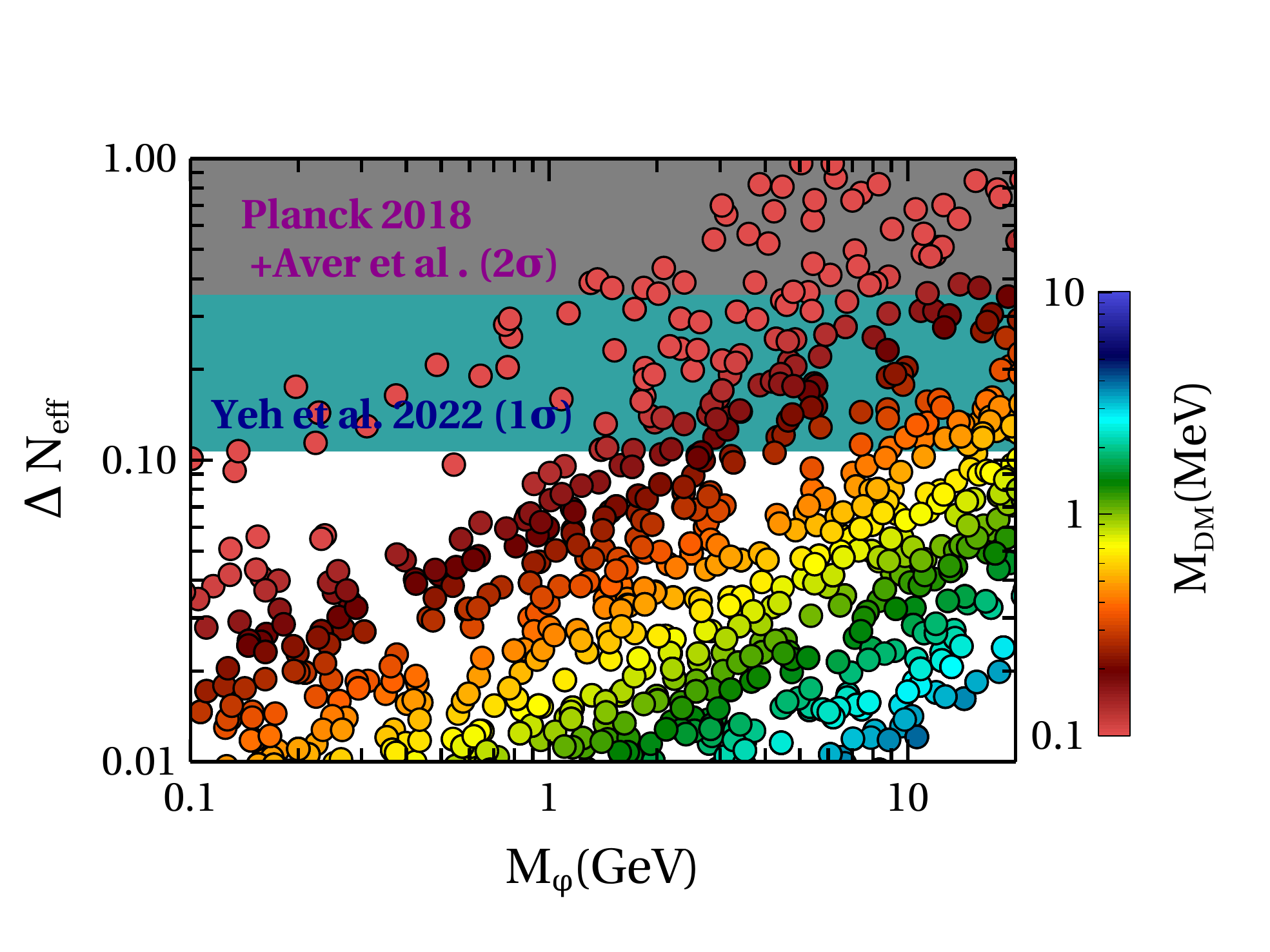}
    \caption{\it Variation of $\Delta N_{\rm eff}$ with $M_{\varphi}$ for
    $\mu_{\varphi}/M_{\varphi}=0.1 $, $\lambda_{\varphi H}=10^{-4}$ and $y_1=10^{-12}$
    where the color gradient represents the range of DM mass  
    in scenario-I. 
    The current upper limits on $\Delta N_{\rm eff}$ from Yeh et al.\cite{Yeh:2022heq} and         Planck 2018\cite{Planck:2018vyg} cobined with Aver et al.\cite{Aver:2015iza} are shown for comparison.Note that the cyan region below the grey band extends upto $\Delta N_{\rm eff}=1.00$ in this plot.}
    \label{fig:neff}
\end{figure}

Finally, we calculate the $\Delta \rm N_{\rm eff}$ for different
values of the model parameters in scenario-I and displayed our 
findings in Fig.\ref{fig:neff}. We present $\Delta \rm N_{\rm eff}$ as a 
function of $M_{\varphi}$ and the color gradient represents the 
range of $M_{\rm DM}$ allowed by observed DM relic density \cite{Planck:2018vyg}. In the figure, we show 
different existing exclusion bounds   
on $\Delta \rm N_{\rm eff}$ depicted by different coloured patches.
We notice that a decrease in $M_{\rm DM}$ yields a increase in
$\Delta \rm N_{\rm eff}$ also. This is easily understood from the fact that lower 
value of $M_{\rm DM}$ requires higher value of $Y_{\chi}$ to satisfy the 
observed relic density.
As we analyzed earlier, $Y_{\chi}$ is governed by $Y_{\varphi}$ and higher value of $Y_{\chi}$ corresponds to 
higher value of $Y_{\varphi}$.
Thus for a higher value of $Y_\varphi$, more energy gets transferred to the neutrino sector, leading to the higher value of $\Delta \rm N_{\rm eff}$.
We also notice that higher values of $M_{\varphi}$ corresponds to the points yielding higher value of  $\Delta \rm N_{\rm eff}$.
This is also understandable  as higher value of $M_{\varphi}$ leads to higher value of $Y_{\varphi}$  resulting higher value of  $\Delta \rm N_{\rm eff}$ for the same reason discussed above.
{In the same plot, we present the current upper 
limits on $\Delta N_{\rm eff}$ . 
However, it is very crucial to note that applying the Planck 2018 bound, 
$N_{\rm eff}=2.99^{+0.33}_{-0.34}$ to our case would be incorrect as 
that bound was derived assuming $N_{\rm eff}$ takes the same value during 
BBN ($N_{\rm eff}^{\rm BBN}$) and before CMB ($N_{\rm eff}^{\rm CMB}$). 
In our scenario, $N_{\rm eff}$ takes the standard value ($=3.046$) 
during BBN and increases due to entropy injection between BBN and CMB epoch. 
A recent analysis conducted by Yeh et al. \cite{Yeh:2022heq}, which aimed to 
investigate entropy injection between BBN and CMB epoch, pointed out an upper limit on 
${\rm DN}_{\rm eff}\equiv N_{\rm eff}^{\rm CMB}- N_{\rm eff}^{\rm BBN}<0.27(\text{ in}~ 2\sigma \text{ limit).}$
For our model,
we translate this bound\footnote{For our case, $\Delta {\rm N}_{\rm eff}^{\rm CMB} \equiv {\rm DN}_{\rm eff} $ considering $N_{\rm eff}^{\rm BBN}=3.046$.} on $\Delta {\rm N}_{\rm eff}^{\rm CMB}$ within $1\sigma $
by the cyan color in Fig.\ref{fig:neff} which represents the exclusion region.
There also exists a BBN independent bound on 
$\Delta \rm N_{\rm eff}$ from Planck 2018 by varying both $N_{\rm eff}$ and 
Helium fraction ($Y_p$) which is $N_{\rm eff}=2.89^{+0.63}_{-0.57}$ in $2\sigma$ limit. 
However, combining  this bound from Planck 2018 data with local measurement from Aver et al.\cite{Aver:2015iza}, slightly 
improves the bound as $N_{\rm eff}=2.99^{+0.43}_{-0.40}$ in $2\sigma$ limit 
\cite{Planck:2018vyg}. 
This bound is also portrayed in Fig.\ref{fig:neff},
shown by the grey region that excludes ($\Delta {\rm  N}_{\rm eff}\gtrsim 0.37$).
Therefore, the most stringent bound applicable for our analysis is $\Delta {\rm  N}_{\rm eff}\lesssim 0.11(\text{ in}~ 1\sigma \text{ limit)}$ \cite{Yeh:2022heq}, which excludes DM 
mass below a few hundred keV.
} 
The future generation experiments like SPT-3G \cite{SPT-3G:2019sok} and CMB-S4 \cite{CMB-S4:2016ple} may probe heavier DM mass.
The allowed parameter space from the constraints of $\Delta \rm N_{\rm eff}$ is also consistent with bound on free streaming length of DM coming from Lyman-$\alpha$ forest\cite{Boyarsky:2008xj,Decant:2021mhj}.


\section{Conclusion}
\label{sec:conc}
In this work we have proposed a minimal extension of the Type-I seesaw model 
with a complex scalar singlet($\varphi$) and a singlet Dirac fermion ($\chi$). 
To ensure the stability of the lightest dark sector particle, an additional 
$\mathcal{Z}_3$ symmetry has been imposed under which $\varphi$ and $\chi$ 
transform non-trivially while the rest of SM particles and three RHNs 
transform trivially. 

Mass spectrum of the dark sector particles are such that the Dirac 
fermion $\chi$ is the lightest particle and plays the role of DM, 
while the singlet scalar $\varphi$ is the next to lightest particle. 
The DM with its tiny coupling with SM bath can only be produced 
from the late time decay of $\varphi$ and obtains its abundance. 
On the other hand $\varphi$ remains in thermal bath due to its strong 
self coupling and after its freeze out it decays to DM 
and active neutrinos. Depending on the thermal history of $\varphi$, 
we have divided the analysis into two scenarios.
In the first Scenario (I), $\varphi$ gains its number density through 
freeze out mechanism via the number changing strong self-interactions 
within the dark sector whereas, in the second Scenario (II) $\varphi$ freezes 
out via the 
SM Higgs portal coupling to SM particles .
 The RHNs($N_{1,2,3}$) which are responsible for generating 
light neutrino masses and mixing angles by type-I seesaw model,
are sufficiently heavy ($M_{N_{1,2,3}}\gg T_{\rm RH}$) such that their number densities do not contribute to DM relic.
However, the presence of RHNs in the particle content allows an effective interaction between $\varphi$,~$\chi$ and active neutrinos($\nu$) 
which leads to extra neutrino production from the late time decay of $\varphi$. 
To track the abundances of $\varphi$ and $\chi$ we have solved two coupled Boltzmann equations.
We have first checked the effects of different model parameters on the relic 
density of DM by solving those  Boltzmann equations and identifying the parameter space giving correct relic density in both scenarios (I $\&$ II).
Apart from producing the right amount of DM relic, the late time decay of $\varphi$ makes
significant impact on the total radiation energy density at the time of CMB formation which is parameterized as $\Delta{\rm N_{\rm eff}}$. 
To compute $\Delta{\rm N_{\rm eff}}$ we have evaluated the extra radiation energy density injected into light neutrino bath from $\varphi$ by solving the required Boltzmann equation.
In scenario-I DM mass up to a few hundred keV is excluded from the present $1\sigma$ limit on $\Delta{\rm N_{\rm eff}}$ from the bound derived by Yeh et al.\cite{Yeh:2022heq}.
The future generation experiments like SPT-3G, CMB-S4
will be sensitive enough to test  heavier DM mass.
However, in scenario-II where the abundance of the mother particle ($\varphi$) is 
suppressed due to sizable interactions with SM bath, we have found that the 
entropy injection 
is insensitive to the bounds on $\Delta{\rm N_{\rm eff}}$ coming from present and 
future-generation experiments.
Thus in this paper we have explicitly shown an alternative way of 
probing FIMP dark matter from the precise measurement of $\Delta{\rm N_{\rm eff}}$ 
even when the mother particles do not have sizable interactions with SM bath which 
is otherwise absent in literature. Consequently, we are expecting 
some very exciting results from next generation CMB experiments, like SPT-3G and 
CMB-S4 which can shed some light on various dark sector models, like the one 
discussed in this paper.
\section*{Acknowledgement} 
SJ and PG thanks D. Nanda for the helpful discussions during this project.
The authors would like to thank Abhijit Kumar Saha, Sougata Ganguly
and Deep Ghosh for useful discussion and comments.
SJ is funded by CSIR, Government of India, under the NET JRF fellowship
scheme with Award file No. 09/080(1172)/2020-EMR-I.

\appendix
\section{Theoretical constraints}
\label{sec:tc}
\section*{Stability}
\label{sec:stab}
The scalar potential is bounded from below when the quartic couplings of the scalar potential satisfy these co-positivity conditions\cite{Kannike:2012pe}:
\begin{align}
\lambda_H \geq 0, ~~\lambda_\varphi \geq 0, ~~
\lambda_{\varphi H} +2 \sqrt{\lambda_\varphi \lambda_H} \geq 0 ~.
\end{align}

The estimation of the lifetime of the desired
the stable vacuum which essentially puts an upper bound on the trilinear dark coupling as \cite{Hektor:2019ote} 
\begin{eqnarray}
\mu_\varphi/M_\varphi < 2 \sqrt{\lambda_\varphi}.
\end{eqnarray}
\section*{Perturbative unitarity}
The tree-level unitarity of the theory, coming from all possible $2\to2$ scattering  amplitudes will form the $S$ matrix and constrain the quartic couplings of the scalar potential\cite{Horejsi:2005da}. The eigenvalues of the $S$ matrix are bounded from above as\cite{Bhattacharya:2017fid}: 
\begin{eqnarray}
\nonumber
 |\lambda_H| \leq 4\pi,~~|\lambda_{\varphi H}| \leq 8\pi,~~|\lambda_\varphi| \leq 4\pi,  \\
 |2\lambda_\varphi+3\lambda_H \pm \sqrt{2\lambda_{\varphi H}^2+(2\lambda_\varphi-3\lambda_H)^2}| \leq 8\pi ~.
 \end{eqnarray}
 The quartic  and Yukawa couplings of the interaction Lagrangian should also obey following inequality equations to maintain perturbativity\cite{Lerner:2009xg}:
\begin{align}
 |\lambda_H| \lesssim  \frac{2\pi}{3},
 |\lambda_{\varphi}| \lesssim \pi,~|\lambda_{\varphi H}| \lesssim 4\pi,  \nonumber \\
 {\rm and}~~~ |y_{\varphi N}| < \sqrt{4\pi} ~.
\end{align}
\section{Constraint from Higgs invisible decay}
\label{sec:higgs}
The dark complex scalar, $\varphi$ is very weakly coupled with SM Higgs via the Higgs portal interaction. The late time decay of $\varphi$ decides both the relic abundance of DM and the contribution to the ${\rm \Delta N_{eff}}$ which require a light scalar mass of the order of  
MeV-few GeV which is well below $M_h/2$ (will be discussed in the next section). In that case, Higgs can  decay to the dark scalar, $\varphi$, and contribute to Higgs's invisible decay width. The Higgs invisible decay width is given by \\
\begin{equation}
\Gamma_{h\rightarrow \varphi \varphi*}= \frac{(\lambda_{\varphi H} v)^2}{16 \pi M_{h}}\sqrt{1-\frac{4 M_{\varphi}^2}{M_h^2}} ~,
\end{equation}
where $M_h=125.06$ GeV and $v=246$ GeV. The current analysis of the CMS collaboration \cite{CMS:2018yfx} at LHC puts a strong constraint on the Higgs invisible decay in the following form
\begin{equation}
{\rm BR^{inv}}= \frac{\Gamma^{\rm inv}_h}{\Gamma^{\rm inv}_h + \Gamma^{\rm SM}_h}<11 \%~ ~,
\end{equation}
where $\Gamma^{SM}_{h}=4$ MeV. If $M_h < 2 M_{\varphi}$ then this decay is absent. However this bound can be further improved as discussed in ref.\cite{Biekotter:2022ckj}. In this work the Higgs invisible constraint only applicable for scenario-II where we require relatively large $\lambda_{\varphi H}$.

\section{  $3\varphi \rightarrow 2\varphi$}
\label{sec:3DMto2DM}
\begin{figure}[h]
\centering
\includegraphics[scale=0.45]{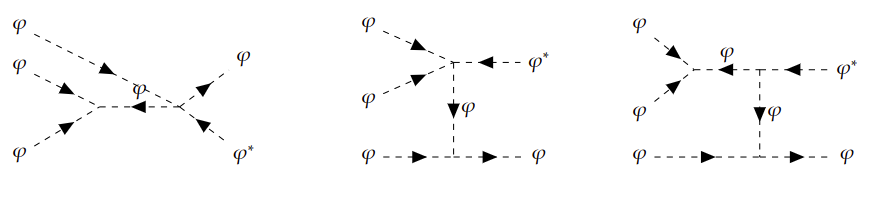} 
 \caption{\it Feynman diagrams for $\varphi~\varphi~\varphi \to \varphi~\varphi^*$ number changing processes. Note that for each t-channel, there is an u-channel diagram.}
\label{fig:FD0}
\end{figure}

\begin{figure}[h]
\centering
\includegraphics[scale=0.45]{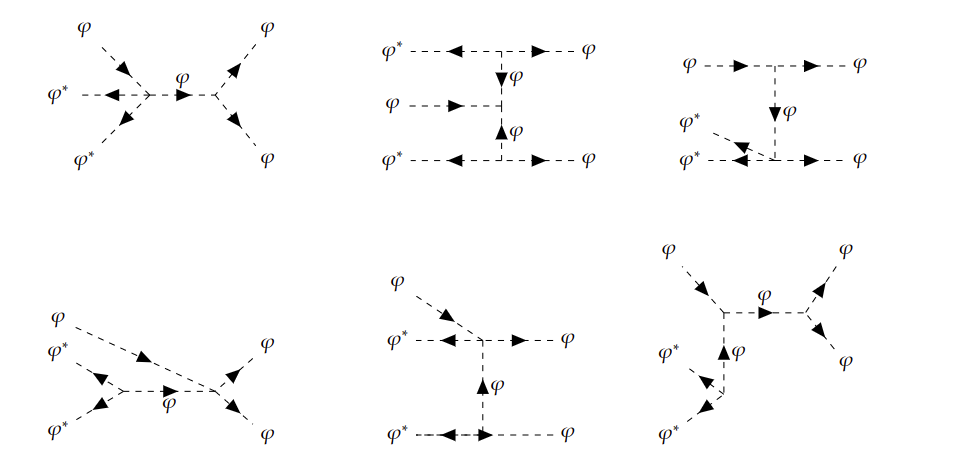} 
 \caption{\it Feynman diagrams for $\varphi~\varphi^*~\varphi^* \to \varphi~\varphi$ number changing processes. Note that for each t-channel, there is an u-channel diagram.}
\label{fig:FD1}
\end{figure}

In our setup $3\varphi \to 2\varphi$ number changing processes in dark sector occur through $\varphi~\varphi~\varphi \to \varphi~\varphi^*$, $\varphi~\varphi^*~\varphi^* \to \varphi ~\varphi $ and their conjugate processes i.e. $\varphi^*~\varphi^*~\varphi^* \to \varphi^*~\varphi$,~ $\varphi^*~\varphi^*~\varphi \to \varphi^* ~\varphi^*$ respectively.
Some of these processes are mediated by $\varphi$ only and the rest are mediated by both $\varphi$ and $h$.
However, for light $M_\varphi$($ \lesssim
\mathcal{O}({\rm GeV})$), h-mediated diagrams are heavily suppressed  due to heavy propagator suppression and small Higgs portal coupling, $\lambda_{\varphi H}$. Therefore for simplicity, one can ignore the Higgs-mediated diagrams. All the $\varphi$ mediated Feynman diagrams for $\varphi~\varphi~\varphi \to \varphi~\varphi^*$ and $\varphi~\varphi^*~\varphi^* \to \varphi~\varphi$ processes are shown in Fig.\ref{fig:FD0} and Fig.\ref{fig:FD1} respectively.  

The amplitude for $\varphi~\varphi~\varphi \to \varphi~\varphi^*$ number changing scattering processes is given by
\begin{eqnarray}
\mathcal{M}_{\varphi\varphi\varphi \to \varphi\varphi^*}&=& \mathcal{M}_1 + \mathcal{M}_2^t + \mathcal{M}_2^u + \mathcal{M}_3^t+\mathcal{M}_3^u\nonumber \\
&=& \Big[ \frac{ 4 \mu_\varphi \lambda_\varphi}{\big(s-M_{\varphi}^{2}\big)}+\frac{4 \mu_\varphi \lambda_\varphi}{\big(t-M_{\varphi}^{2}\big)}+\frac{4 \mu_\varphi \lambda_\varphi}{\big(u-M_{\varphi}^{2}\big)}+\frac{\mu_\varphi^3 }{\big(s-M_{\varphi}^{2}\big) \big(t-M_{\varphi}^2\big)}+\frac{\mu_\varphi^3 }{\big(s-M_{\varphi}^{2}\big) \big(u-M_{\varphi}^2\big)} \Big] . \nonumber \\
\end{eqnarray}

And the amplitude for $\varphi~\varphi^*~\varphi^* \to \varphi~\varphi$ number changing scattering processes is given by
\begin{eqnarray}
\mathcal{M}_{\varphi^*\varphi^*\varphi \to \varphi\varphi}&=& \mathcal{M}_1 + \mathcal{M}_2^t + \mathcal{M}_2^u + \mathcal{M}_3^t+\mathcal{M}_3^u+\mathcal{M}_4 +\mathcal{M}_5^t+\mathcal{M}_5^u++\mathcal{M}_6^t+\mathcal{M}_6^u\nonumber \\
&=& \Big[ \frac{4\mu_\varphi \lambda_\varphi}{\big(s-M_\varphi^2\big)} + \frac{\mu_\varphi^3}{\big( t- M_\varphi^2\big)^2} + \frac{\mu_\varphi^3}{\big( u- M_\varphi^2\big)^2}+\frac{4\mu_\varphi \lambda_\varphi}{\big(t-M_\varphi^2\big)}+\frac{4\mu_\varphi \lambda_\varphi}{\big(u-M_\varphi^2\big)} +\frac{4\mu_\varphi \lambda_\varphi}{\big(s-M_\varphi^2\big)} \nonumber \\
&&~~~~~~~~~~ +\frac{4\mu_\varphi \lambda_\varphi}{\big(t-M_\varphi^2\big)}+\frac{4\mu_\varphi \lambda_\varphi}{\big(u-M_\varphi^2\big)}+\frac{\mu_\varphi^3}{\big(t-M_\varphi^2\big)\big(s-M_\varphi^2\big)}+\frac{\mu_\varphi^3}{\big(u-M_\varphi^2\big)\big(s-M_\varphi^2\big)}\Big] . \nonumber \\
\end{eqnarray}

The total thermal averaged cross section for $3\varphi \to 2\varphi$ number changing processes can be expressed using non-relativistic approximation as\cite{Pierre:2018man}:
\begin{eqnarray}
\langle \sigma v^2 \rangle _{3\varphi \to 2\varphi}&=& \langle \sigma v^2 \rangle _{\varphi \varphi\varphi \to \varphi \varphi^*}+\langle \sigma v^2 \rangle _{\varphi \varphi^*\varphi^* \to \varphi \varphi} \nonumber \\
&\approx&\frac{\sqrt{5}}{192\pi M_\varphi^3} \Big(|\mathcal{M}_{\varphi \varphi\varphi \to \varphi \varphi^*}|^2+|\mathcal{M}_{\varphi^* \varphi^*\varphi^* \to \varphi \varphi^*}|^2  \Big) \nonumber \\ &&~~+\frac{\sqrt{5}}{192\pi M_\varphi^3} \Big(|\mathcal{M}_{\varphi \varphi^*\varphi^* \to \varphi \varphi}|^2+|\mathcal{M}_{\varphi^* \varphi\varphi \to \varphi^* \varphi^*}|^2  \Big) \nonumber \\
&=& \frac{\sqrt{5}}{192\pi M_\varphi^3} \Big(2|\mathcal{M}_{\varphi \varphi\varphi \to \varphi \varphi^*}|^2+2|\mathcal{M}_{\varphi \varphi^*\varphi^* \to \varphi \varphi}|^2  \Big),
\end{eqnarray}
where  $|\mathcal{M}_{\varphi \varphi \varphi \to \varphi \varphi^*}|^2=|\mathcal{M}_{\varphi^* \varphi^*\varphi^* \to \varphi \varphi^*}|^2$ and $|\mathcal{M}_{\varphi \varphi^*\varphi^* \to \varphi \varphi}|^2=|\mathcal{M}_{\varphi^* \varphi\varphi \to \varphi^* \varphi^*}|^2$.

\section{ $2\varphi \to 2 {\rm~SM}$ and   $\varphi~{\rm SM} \to \varphi ~{\rm SM}$}
\label{sec:2DMto2SM}
\begin{figure}[h]
\centering
\includegraphics[scale=0.45]{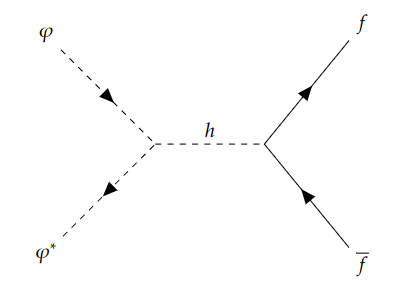} 
 \caption{\it Feynman Diagrams for $\varphi~\varphi^* \to f\overline{f}$ where $f$ stands for SM fermions excluding top quark.}
\label{fig:FD2}
\end{figure}

There is another type of number-changing process between the dark sector, $\varphi$, and the visible sector, SM where two dark scalar $\varphi$ annihilates into two SM particles via $h$ mediated diagram. Note that our analysis mostly focuses on the light-dark scalar
with mass up to a few GeV. Therefore $\varphi$ can only annihilate into light fermion pairs. The Feynman diagrams of corresponding number-changing processes are shown in Fig.\ref{fig:FD2}.

\noindent The thermal averaged cross-section for $2\varphi \to 2 {\rm SM}$ number changing process is given by:
\begin{eqnarray}
\langle \sigma v \rangle_{2\varphi \to 2 {\rm SM}} &=& \sum_{f} \langle \sigma v \rangle_{\varphi\varphi^* \to f\overline{f}} \nonumber \\
&=& \sum_{f}  \frac{x}{16 T M_\varphi^4 K_2(x)^2} \int_{4 M_{\varphi}^2}^{\infty}  \big( \sigma v\big)_{\varphi\varphi^* \to f \overline{f}} K_1\big(\frac{\sqrt{s}}{T}\big) s \sqrt{s-4M_\varphi^2} ~ds \nonumber \\
\end{eqnarray}
where $x=\frac{M_\varphi}{T}$ and $\big( \sigma v \big)_{\varphi\varphi^* \to f \overline{f}}$  can be written as:
\begin{eqnarray}
(\sigma v)_{\varphi \varphi \rightarrow f \overline f}&=&\Big(\frac{1}{4\pi s \sqrt s} \frac{N_c\lambda_{\varphi H}^2 m_f^2}{(s-m_h^2)^2+m_h^2 \Gamma_h^2}(s-4m_f^2)^\frac{3}{2} \Big) \Theta(M_\varphi - m_f).
\label{eq:annihilationSM}
\end{eqnarray}
In the above expression $N_c=1$ for leptons and $N_c=3$ for quarks. However for scenario-II we considered $M_\varphi$ upto $100$ GeV where interactions having Higgs mediated gauge boson final states are contributing to $2\varphi \to 2{\rm SM}$ processes. 

\begin{figure}[h]
\centering
\includegraphics[scale=0.45]{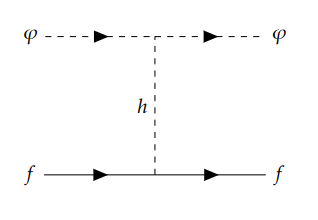} 
 \caption{\it Feynman Diagrams for $\varphi~f \to \varphi f$ where $f$ stands for SM fermion excluding top quark.}
\label{fig:FD3}
\end{figure}

The scattering between DM and SM,
$\varphi ~f \to \varphi ~f$
 is also important for our discussion which is required for
analysing the kinetic equilibrium of the DM in early universe. The Feynman diagram for the scattering between DM and SM fermions are shown in Fig.\ref{fig:FD3}.

\noindent The thermal averaged scattering cross-section between DM and SM is followed by:
\begin{eqnarray}
\langle \sigma v \rangle_{\varphi ~f\to \varphi~ f} &=&  \sum_{f}  \Big \langle \sigma v \rangle_{\varphi f \to \varphi  f} + \langle \sigma v \rangle_{\varphi^* f \to \varphi^*  f} \Big)= 2 \sum_{f}  \langle \sigma v \rangle_{\varphi f \to \varphi  f} \nonumber \\
&=& \sum_{f}  \frac{x}{16 T M_\varphi^2 m_f^2  K_2(M_\varphi/T) K_2(m_f/T)}    \nonumber \\
&&~~~~~~\times \int_{\big( M_{\varphi}+m_f \big)^2}^{\infty}  \big( \sigma v\big)_{\varphi f \to \varphi f } K_1\big(\frac{\sqrt{s}}{T}\big) s \sqrt{s-\big(M_\varphi+m_f \big)^2} ~ds ~,
\end{eqnarray}
where $x=\frac{M_\varphi}{T}$ and the scattering cross-section, $\big( \sigma v\big)_{\varphi f \to \varphi f }$ is given by,
\begin{eqnarray}
\big( \sigma v\big)_{\varphi f \to \varphi f } &=& \frac{1}{4\pi s\sqrt{s}} \frac{1}{2\sqrt{s}} \sqrt{\big[s-\big(M_\varphi+m_f \big)^2\big]\big[s-\big(M_\varphi-m_f \big)^2\big]} \nonumber \\
&&~~~~~~~~~~\times \Big[ -2 (t-4M_\varphi^2) \Big(  \frac{\lambda_{\varphi H} v}{t-m_h^2}\Big)^2  \Big]~.
\end{eqnarray}

\section{Condition for Kinetic equilibrium}
\label{sec:KEQ}
Following the discussion in sec.\ref{sec:dynamics} the Kinetic equilibrium is ensured by the following inequality,
\begin{equation}
\Gamma_{ [\varphi~f \to \varphi~f]}(T)\gtrsim \mathcal{H}(T).
\end{equation} 
As shown in Fig.\ref{fig:FD3} $\varphi$ undergoes elastic scattering processes with SM fermions.
The interaction rate is given by,
\begin{equation}
\Gamma_{[\varphi~f \to \varphi~f]}(T)= n_f^{eq}(T) \langle \sigma v \rangle_{\varphi f\to \varphi f},
\end{equation}
where, $ n_f^{eq}(T)$ is the equlibrium number densities of fermions and given by,
\begin{equation}
 n_f^{eq}(T)=\frac{g_f}{(2\pi)^{3/2}}4 \pi m_f^2 T~ K_2\left(\frac{m_f}{T} \right)
\end{equation}

\begin{figure}[tbh]
\centering
\includegraphics[scale=0.4]{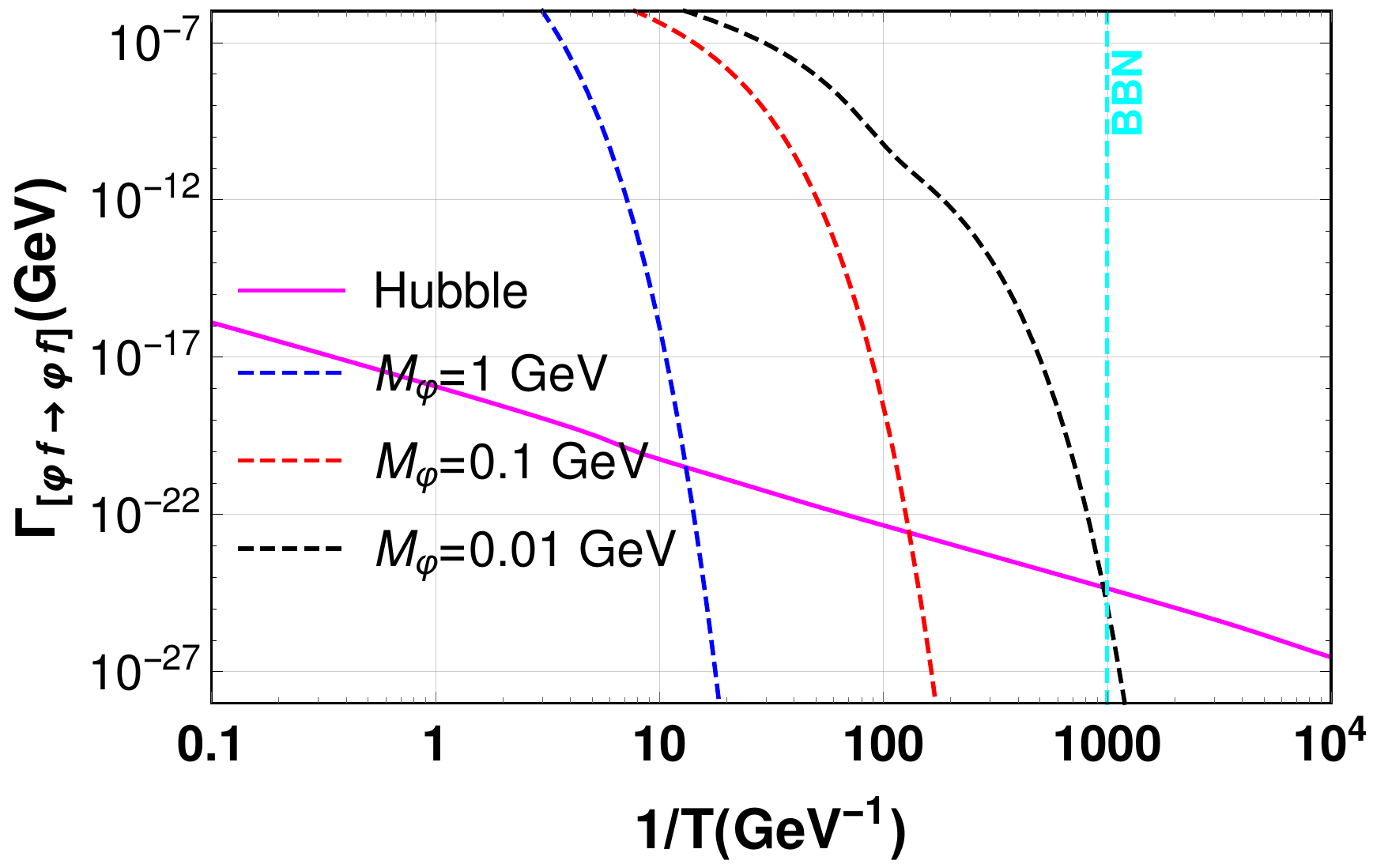} 
 \caption{\it Evoluation of $\Gamma_{\rm [\varphi~f \to \varphi~f]}$ with $1/T$ keeping $\lambda_{\varphi H}=10^{-4},~\mu_{\varphi}/M_{\varphi}=0.1$ for three different mass of $\varphi$, $M_{\varphi}=1,0.1,0.01$ GeV depicted by blue,red and black dashed line respectively. The solid magenta line signifies $\mathcal{H}(T)$.}
\label{fig:KE}
\end{figure}
In Fig.\ref{fig:KE} we display the evolution of elastic scattering rate $\Gamma_{[\varphi~f\to \varphi~f]}$ with $1/T$ for different values of $M_{\varphi}=1,0.1,0.01$ GeV depicted by blue,red and black dashed line respectively.
In the same plot we display the expansion rate $\mathcal{H}$ for comparision, shown by the solid magenta line.
We notice that the condition for kinetic equilibrium doesn't hold after $T <  M_{\varphi}/10$.
In this work we took the lowest value of $M_{\varphi}=0.1$ GeV and the kinetic equilibrium breaks long before BBN ($T\sim 1$ MeV).
Note that the kinetic equlibrium can be possible even after BBN for $M_\varphi \lesssim 0.01$ GeV which can give rise to observable imprints \cite{Ho:2012ug,Boehm:2012gr}.
For example ref.\cite{Boehm:2012gr} has shown that if particle with $\mathcal{O}$(MeV) mass remains in thermal equilibrium with neutrino bath after neutrino decoupling leads to increament in $N_{\rm eff}$. 
However, to avoid such complications, we restrict the parameter space in our work with lowest mass of $\varphi$ as $0.1$  GeV such that the kinetic decoupling happens long before BBN ($T\sim 1$ MeV).
\bibliography{ref}
\end{document}